\documentclass[twocolumn]{article}

\usepackage[english]{babel}

\usepackage[top=1.5cm,bottom=1.5cm,left=1.5cm,right=1.5cm,marginparwidth=1.75cm]{geometry}

\usepackage{amsmath}
\usepackage{graphicx}
\usepackage[colorlinks=true, allcolors=blue]{hyperref}

\usepackage{amssymb}
\usepackage{tabularx}
\usepackage[table,xcdraw]{xcolor}

\usepackage{siunitx}

\usepackage{textgreek}
\usepackage{comment}
\usepackage{authblk}

\usepackage{enumitem}
\usepackage{url}
\usepackage{natbib}
\usepackage[nohyperlinks]{acronym}

\title{\vspace{-1cm}GENESIS: Co-location of Geodetic Techniques in Space}

\author[1]{Pac\^ome Delva} 
\affil[1]{SYRTE, Observatoire de Paris-PSL, Sorbonne Université, CNRS UMR8630, LNE, 61 avenue de l'Observatoire, 75014 Paris, France}
\author[2]{Zuheir Altamimi} 
\affil[2]{Université de Paris Cité, Institut de physique du globe de Paris, CNRS, IGN, F-75005 Paris, France}
\author[3,4]{Alejandro Blazquez} 
\affil{LEGOS, Université de Toulouse (CNES, CNRS, IRD, UPS), 14 avenue Edouard Belin, 31401 Toulouse, France}
\affil[4]{Centre National d’Etudes Spatiales, 18 avenue Edouard Belin, 31401 Toulouse, France}
\author[5]{Mathis Blossfeld} 
\affil[5]{DGFI, Technische Universität München, Arcisstraße 21 80333 München, Germany}
\author[6]{Johannes B\"ohm} 
\affil[6]{Technische Universit{\"a}t Wien, Wiedner Hauptstraße 8-10, 1040 Vienna, Austria}
\author[1]{Pascal Bonnefond}
\author[7]{Jean-Paul Boy} 
\affil[7]{Institut Terre \& Environnement de Strasbourg, Université de Strasbourg, CNRS UMR7063, 5 rue René Descartes, 67084 Strasbourg, France}
\author[4,8]{Sean Bruinsma}
\affil[8]{GET, Université de Toulouse (CNES, CNRS, IRD, UPS), 14 avenue Edouard Belin, 31401 Toulouse, France}
\author[9]{Grzegorz Bury} 
\affil[9]{Institute of Geodesy and Geoinformatics, Wroclaw University of Environmental and Life Sciences, Norwida 25, 50-375 Wroclaw, Poland}
\author[1]{Miltiadis Chatzinikos}
\author[4,8]{Alexandre Couhert}
\author[10]{Cl\'ement Courde} 
\affil[10]{Université Côte d'Azur, CNRS, Observatoire de la Côte d'Azur, IRD, Géoazur, 2130 Route de l'Observatoire 06460 Caussols, France}
\author[11]{Rolf Dach} 
\affil[11]{Astronomical Institute of the University of Bern, Sidlerstrasse 5, 3012 Bern, Switzerland}
\author[12]{V\'eronique Dehant} 
\affil[12]{Royal Observatory of Belgium, Ringlaan 3, BE-1180, Brussels, Belgium}
\author[13]{Simone Dell’Agnello} 
\affil[13]{National Institute for Nuclear Physics -- Frascati National Labs (INFM-LNF), via E. Fermi 54, Frascati (Rome), 00044, Italy}
\author[14]{Gunnar Elgered} 
\affil[14]{Chalmers University of Technology, Onsala Space Observatory, SE 439 92 Onsala, Sweden}
\author[15]{Werner Enderle} 
\affil[15]{European Space Operations Center, ESA/ESOC, 64293 Darmstadt, Germany}
\author[8]{Pierre Exertier} 
\author[16]{Susanne Glaser} 
\affil[16]{German Research Centre for Geosciences (GFZ), Telegrafenberg, 14473 Potsdam, Germany}
\author[14]{R\"udiger Haas}
\author[16]{Wen Huang}
\author[17]{Urs Hugentobler} 
\affil[17]{Institut für Astronomische und Physikalische Geodäsie, Technische Universität München, 80290 München, Germany}
\author[11]{Adrian J\"aggi}
\author[12]{Ozgur Karatekin}
\author[18]{Frank G. Lemoine} 
\affil[18]{Geodesy \& Geophysics Laboratory, NASA Goddard Space Flight Center, Greenbelt, Maryland 20771, U.S.A.}
\author[1]{Christophe Le Poncin-Lafitte}
\author[16]{Susanne Lunz}
\author[16]{Benjamin M\"annel} 
\author[4,8]{Flavien Mercier}
\author[2]{Laurent M\'etivier} 
\author[3,4]{Beno\^it Meyssignac}
\author[19]{J\"urgen M\"uller} 
\affil[19]{Leibniz University Hannover, Institute of Geodesy, Schneiderberg 50, 30167 Hannover, Germany }
\author[6]{Axel Nothnagel}
\author[4,8]{Felix Perosanz}
\author[20]{Roelof Rietbroek} 
\affil[20]  {ITC Faculty of Geo--information Science and Earth Observation, Department of Water Resources (WRS), Hengelosestraat 99
7514 AE, Enschede, Netherlands}
\author[21]{Markus Rothacher} 
\affil[21]{Institute of Geodesy and Photogrammetry, ETH Zurich, Stefano--Franscini--Platz 5, 8093 Zurich, Switzerland}
\author[12]{Hakan Sert}
\author[9]{Krzysztof Sosnica}
\author[22]{Paride Testani} 
\affil[22]{HE Space Operations B.V. for ESA - European Space Agency, 2200 AG Noordwijk, Netherlands}
\author[23]{Javier Ventura-Traveset} 
\affil[23]{ESA Toulouse, Centre Spatial de Toulouse, 18 Avenue Edouard Belin, 31401 Toulouse Cedex 9, France}
\author[24]{Gilles Wautelet} 
\affil[24]{Laboratory of Planetary and Atmospheric Physcis (LPAP), University of Liège, Allée du Six Août, 19C 4000 Liège, Belgium}
\author[9]{Radoslaw Zajdel}

\begin{document}

\maketitle

\abstract{Improving and homogenizing time and space reference systems on Earth and, more directly, realizing the \ac{TRF} with an accuracy of \SI{1}{mm} and a long-term stability of \SI{0.1}{mm/year} are relevant for many scientific and societal endeavors. The knowledge of the \ac{TRF} is fundamental for Earth and navigation sciences. For instance, quantifying sea level change strongly depends on an accurate determination of the geocenter motion but also of the positions of continental and island reference stations, such as those located at tide gauges, as well as the ground stations of tracking networks. Also, numerous applications in geophysics require absolute millimeter precision from the reference frame, as for example monitoring tectonic motion or crustal deformation for predicting natural hazards. The \ac{TRF} accuracy to be achieved represents the consensus of various authorities, including the International Association of Geodesy (IAG), which has enunciated geodesy requirements for Earth sciences. Moreover, the United Nations Resolution 69/266 states that the full societal benefits in developing satellite missions for positioning and remote sensing of the Earth are realized only if they are referenced to a common global geodetic reference frame at the national, regional and global levels.

Today we are still far from these ambitious accuracy and stability goals for the realization of the \ac{TRF}. However, a combination and co-location of all four space geodetic techniques on one satellite platform can significantly contribute to achieving these goals. This is the purpose of the GENESIS mission, proposed as a component of the FutureNAV program of the European Space Agency. The GENESIS platform will be a dynamic space geodetic observatory carrying all the geodetic instruments referenced to one another through carefully calibrated space ties. The co-location of the techniques in space will solve the inconsistencies and biases between the different geodetic techniques in order to reach the \ac{TRF} accuracy and stability goals endorsed by the various international authorities and the scientific community.

The purpose of this white paper is to review the state-of-the-art and explain the benefits of the GENESIS mission in Earth sciences, navigation sciences and metrology. This paper has been written and supported by a large community of scientists from many countries and working in several different fields of science, ranging from geophysics and geodesy to time and frequency metrology, navigation and positioning. As it is explained throughout this paper, there is a very high scientific consensus that the GENESIS mission would deliver exemplary science and societal benefits across a multidisciplinary range of Navigation and Earth sciences applications, constituting a global infrastructure that is internationally agreed to be strongly desirable.}

\section{Introduction} 
\label{sec:intro}
The GENESIS proposal is dedicated to improving and homogenizing time and space references on Earth and, more directly, to realizing the \ac{TRS} with an accuracy of \SI{1}{mm} and a long-term stability of \SI{0.1}{mm/year}. These numbers are relevant for many scientific and societal endeavors for which a precise realization of the \ac{TRS} and the knowledge of the Earth’s kinematic parameters are crucial. 

Knowledge of \ac{CRF} and of \acf{TRF} is fundamental for orbit computation, in particular for metrological applications such as altimetry (e.g., ocean radar altimetry, laser and radar altimetry over land surfaces and ice sheets, and interferometric Synthetic-Aperture Radar mapping of land surface change), as well as for more precise position determinations of Earth orbiting satellites. For instance, quantifying sea level change or the effects of ice melting using altimetry strongly depends on an accurate determination of the position of continental and island reference stations, such as those located at tide gauges, as well as the ground stations of tracking networks. Also, numerous applications in geophysics require absolute millimeter precision from the reference frame, as for example in the case of monitoring tectonic motion or crustal deformation for predicting natural hazards and inferring climate driven mass changes (non-tidal ocean, ice, atmospheric, and hydrological) from observations of vertical and horizontal displacements of the Earth's surface. 

A stable and accurate reference frame is needed for  robust policy making in light of climate change. The quality of many operational monitoring systems are tied to the accuracy of the underlying reference systems. Reliable evidence based policies, which make use of such operational data, and are expected to become more important in adaptation measures, are therefore directly dependent on the quality of international reference frames. The \ac{TRF} accuracy and stability to be achieved, respectively \SI{1}{mm} and  \SI{0.1}{mm/year}, represent the consensus of various authorities, including the \ac{IAG}, which has enunciated geodesy requirements for Earth science through the \ac{GGOS} initiative \citep[see][]{plag_global_2009}. Hereafter we will refer to these numbers as the \ac{GGOS} accuracy and stability goals.

The General Assembly of the \ac{UN} adopted a resolution on 26 February 2015: ``A global geodetic reference frame for sustainable development'' \citep{UN2015a}. In this resolution the \ac{UN} recognize the importance of ``the investments of Member States in developing satellite missions for positioning and remote sensing of the Earth, supporting a range of scientific endeavors that improve our understanding of the Earth system and underpin decision-making, and [...] that the full societal benefits of these investments are realized only if they are referenced to a common global geodetic reference frame at the national, regional and global levels''. Moreover, in this resolution, the \ac{UN} ``invites Member States to engage in multilateral cooperation that
addresses infrastructure gaps and duplications towards the development of a more sustainable global geodetic reference frame''. In this article, we will explain that the GENESIS mission is proposed precisely to address the geodetic ground infrastructure gaps. The \ac{UN-GGIM} established a working group to develop a global geodetic road map that addresses key elements relating to the development and sustainability of the \ac{GGRF}.

Several space missions in order to reach the \ac{GGOS} accuracy and stability goals were proposed in the past, such as GRASP \citep{barsever2009}, E-GRIP \citep{jetzer_e-grip_2016} and E-GRASP \citep{biancale_e-grasp_2017}. Today, the GENESIS mission is timely considering the large number of long-term scientific undertakings, where many different space data need to be analyzed together, as, for example, for the quantification of the mass-loss in the polar regions, where altimetry missions (ICESat-2, CryoSat-2, Sentinel-6) and gravity field missions (GRACE and GRACE-FO) need to be jointly exploited. The strong statements by international bodies underline that the GENESIS mission is highly needed and timely, being the only proposed mission to cover this topic, worldwide.

A very large and strong scientific community involved in the worldwide networks and data and analysis centers of the four geometrical \ac{IAG} Services supports the GENESIS proposal as part of the FutureNAV program of the \ac{ESA}. Clearly, GENESIS will deliver exploratory results across many Earth's science disciplines, mainly where precise positioning, surface motions and mass movement are critical. The mission is thus supported by an active and broad community, and the downstream science and policy-making users will continuously benefit from the improvements in the \ac{TRF} and of its link to the \ac{CRF}.

\section{Summary of GENESIS science and mission objectives} 
\begin{figure}
\centering
\includegraphics[width=1.0\linewidth]{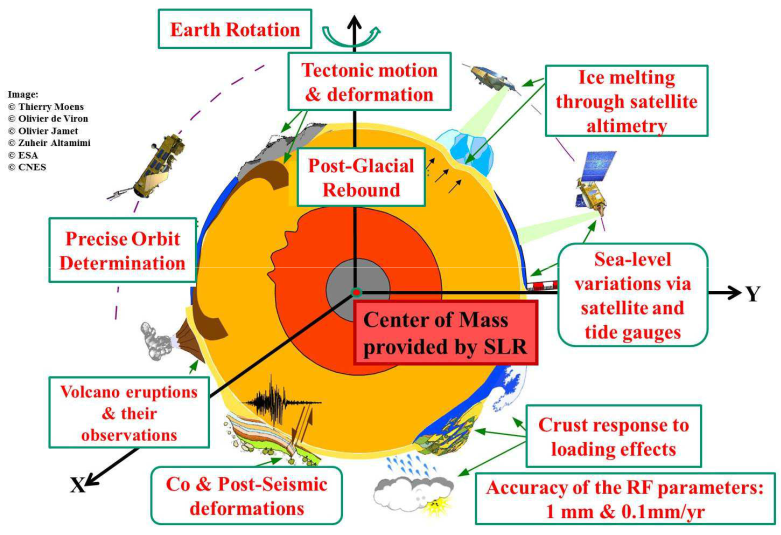}
\caption{\label{fig:apps} GENESIS mission primary goal is a significant improvement of the \acf{ITRF}. The \ac{ITRF} is recognised to be the metrological foundation for all space- and ground-based observations in Earth Science and Navigation, and therefore this mission will potentially have a major impact in a large number of GNSS and Earth Observation applications. This figure shows some of them.}
\end{figure}

The \ac{GGOS} was initiated by the \ac{IAG} with the goal of providing a consistent high-quality \ac{TRF}, crucial for the present day and future science application needs \citep{plag_global_2009, national_academies_of_sciences_thriving_2018, national_academies_of_sciences_evolving_2020}. There exists a very wide spectrum of applications in Navigation and Earth sciences and far beyond that require a more accurate \ac{TRF}, as illustrated in Figure~\ref{fig:apps}. The \ac{TRF} is the indispensable fundamental metrological basis to allow a long-term consistent monitoring of the Earth's system changes.

The GGOS accuracy and stability goals are required to detect the smallest variations in the Earth system components. These requirements are especially driven by the fact that the stability in the present \ac{TRF} of about \SI{0.5}{mm/year} is the most important contribution to the uncertainty in global sea level rise (see Sect.~\ref{sec:AltiSea}) and in many other geophysical processes. The primary goal of the GENESIS mission is, therefore, the establishment of a \ac{TRF} supporting the \ac{GGOS} accuracy and stability goals through the co-location of space geodetic techniques on a single satellite. 

Nowadays, the \ac{TRF} is realized by station coordinates and velocities for a globally distributed set of ground stations using a combination of the four major space geodetic techniques:
\begin{itemize}
    \item \ac{GNSS};
    \item \ac{DORIS}, a radio satellite tracking system;
    \item the \ac{SLR} technique;
    \item and the \ac{VLBI} technique, which normal operation is to record the signals from quasars.
\end{itemize}


In order to develop a unique, consistent and accurate \ac{TRF}, these four techniques are combined and linked together thanks to co-location sites located on the ground, where more than one space geodetic technique is located at the same site. Thereby the local ties, i.e., the vectors connecting the reference points of the individual instruments (\ac{GNSS} and \ac{DORIS} antennas, radio and optical telescopes), must be realized at the \SI{1}{mm} level or better.

Unfortunately, one of the major deficiencies in the realization of a \ac{TRF} originates from the difficulty to accurately measure the local ties between the reference points (intersection of axes of large instruments, phase centres of antennas). A second issue of the reference frame is that each technique has its own systematic effects. Thus, a second deficiency includes the (as yet unknown) systematic effects present in the observations of the individual space geodetic techniques. Finally, a third deficiency is the poor spatial distribution of co-location sites on the globe. 

\begin{figure}
\centering
\includegraphics[width=1.0\linewidth]{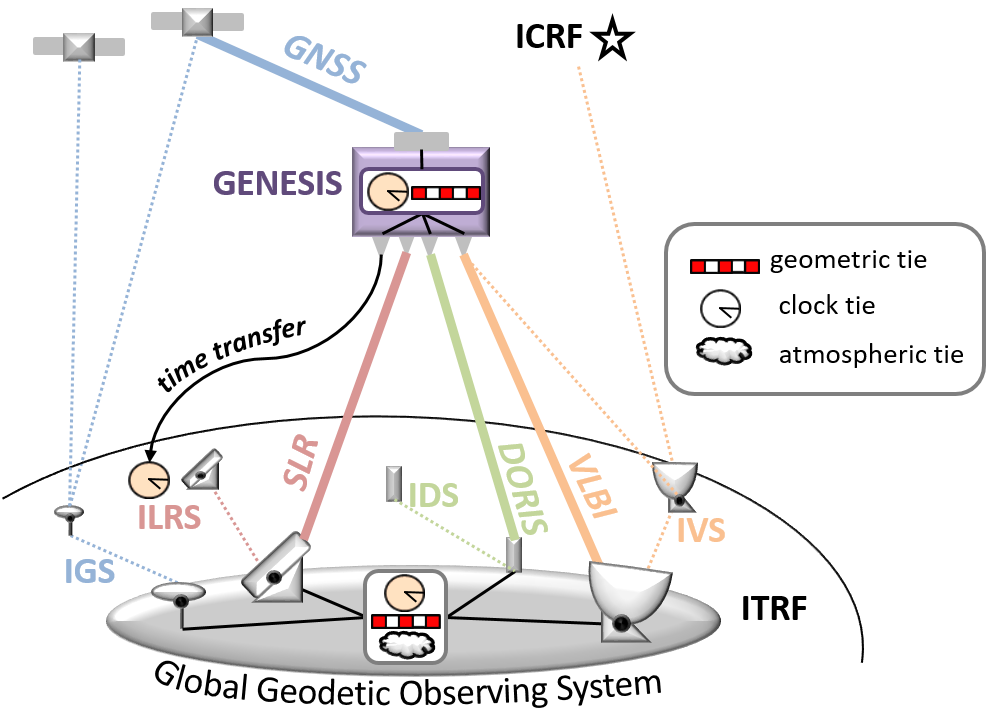}
\caption{\label{fig:concepts} The GENESIS mission will consist of the co-location, for the first time ever, of the four space geodetic techniques (GNSS, SLR, VLBI, and DORIS) aboard a single well-calibrated satellite in \acf{MEO}. This will result in a unique dynamic space geodetic observatory, which combined with the measurements of geodetic co-location sites on the ground, shall allow obtaining a significant improvement of the \acf{ITRF}.}
\end{figure}

In order to improve this situation fundamentally, the GENESIS satellite mission will provide a highly accurate co-location of the four space geodetic techniques in space, on board the satellite with carefully and fully calibrated reference points, as illustrated in Figure~\ref{fig:concepts}. Thus, GENESIS will be a calibrated co-location and reference point, fully complementary to the ground co-location, orbiting in space and connecting all the ground stations to one another. In this way, one can determine all the instrumental biases inherent to the different observing techniques simultaneously. This bias determination is required: (1) to avoid systematic errors, which can result in erroneous interpretations of the differences in the techniques, as well as (2) to transmit the \ac{TRF} via \ac{GNSS} at the precision of a millimeter to any point on the Earth that can then be used for precise positioning and navigation. The payload that will allow us to realize this co-location consists of a \ac{VLBI} transmitter, a \ac{GNSS} and a \ac{DORIS} receiver, a \ac{P-LRR} and a \ac{USO} that will connect all four techniques. 

Secondary goals could be reached by with the addition an \ac{A-LRR} and an accelerometer. The \ac{A-LRR} would allow for a high precision synchronization of the onboard USO with ground clocks through time transfer by laser link from ground stations. This would provide metrology users with a common view, time transfer technique, accurate at intercontinental scales. More specifically, in addition to the benefits of the already present \ac{P-LRR}, the \ac{A-LRR} would allow (1) to perform ground-to-space and ground-to-ground time and frequency transfers with an extended common view compared to the T2L2 and ACES missions, by taking advantage of the higher altitude of the GENESIS satellite, (2) to compare GNSS and laser time transfer techniques with an uncertainty below \SI{100}{\ps} and (3) to accurately monitor the behaviour of the onboard clock for precise orbitography. 

An accelerometer, in combination with a well tested macro model of the satellite's geometry and reflectance, would provide insight in non-conservative forces and their effect on the GENESIS orbit. All the mechanical and electronic properties of the platform must be characterized to within sub-millimeter tolerances. An accelerometer could be used to measure surface accelerations up to $10^{-11}$~\si{m.s^{-2}.Hz^{-1/2}} and would (1) guarantee high-precision orbit determination and mitigating the errors mapping into the modeling of non-conservative forces, (2) allow in-orbit \ac{CoM} determination. These characteristics are important in the determination of station positions and geophysical products such as the geocenter and Earth's orientation parameters. Another role of this accelerometer would be to serve as a position reference for the geodetic instruments on the platform in order to determine the correction of angular motion between each instrument.

The availability of high-precision measurements from the GENESIS mission will fundamentally improve the accuracy and stability of the \ac{TRF} by a factor of 5-10 and will allow us to achieve the \ac{GGOS} requirements. This paper reviews the science applications of the GENESIS mission (see also Figure~\ref{fig:apps}): improvements in the \ac{TRF} geocenter and scale (Sect.~\ref{sec:GeocScale}); improvements in the celestial (inertial) frame and the Earth orientation parameters, reflecting Earth system processes (Sect.~\ref{sec:UnifFrame}); improvements in the knowledge of the low-degree spherical harmonics of the Earth gravity field, complementary to GRACE and GRACE-FO (Sect.~\ref{sec:longwave}); improvements in global to local estimates of sea level change (Sect.~\ref{sec:AltiSea}); improvement in estimates of present day ice mass loss and \ac{GIA} history (Sect.~\ref{sec:IceMass}); improved determination of the Earth’s rheology and the melting history (Sect.~\ref{sec:geo}); improved quantification of surface loads due to the continental water cycle, the atmosphere and the ocean (Sect.~\ref{sec:geo}); improvements in the Earth radiation budget (Sect.~\ref{sec:RadBudget}); improvements of 
the ionospheric and plasmaspheric density (Sect.~\ref{sec:IonoPlasma}); distribution of a high-accuracy reference frame to all \ac{GNSS} users for global geo-referencing at the millimeter-level (Sect.~\ref{sec:GlobPos}); very accurate and consistent antenna phase centre calibrations for all \ac{GNSS} satellites relevant for the terrestrial scale and all positioning applications (Sect.~\ref{sec:GNSS}); millimeter-level precise orbit determination (POD) for altimetric, gravimetric and \ac{GNSS} satellites (Sect.~\ref{sec:PosSat}); intercontinental time transfer at the picosecond level and its use to unify height systems by exploiting the gravitational redshift (Sect.~\ref{sec:RelGeo}). Since reference frames are at the heart of metrology and all monitoring processes, the benefits from GENESIS are truly inter- and transdisciplinary and relevant for societal needs.

Moreover, an \ac{ESA} \ac{CDF} study has confirmed that the GENESIS Mission is feasible within the \ac{ESA} FutureNAV defined program boundaries, with a target launch date in 2027. The conclusion of this study is given in Sect.~\ref{sec:CDF}. This section is followed by a high-level description of laser ranging (Sect.~\ref{sec:laser}) and VLBI transmitter (Sect.~\ref{sec:VLBI}).

\section{Reference frames}
\label{sec:RefFrames0}

\subsection{Importance of reference frames}\label{sec:RefFrames} %

Reference frames provide the necessary absolute basis for the relative-only geodetic measurements. They are indispensable to study the dynamic Earth, and to be able to meaningfully relate changes across space and time.
They are also essential for positioning and navigation in the civil society and for proper georeferencing of geospatial information. The provision of accurate and stable reference frames is one of the major tasks of geodesy.  

The \ac{TRF} is the realization of the \ac{TRS} and is currently provided by precisely determined coordinates and velocities of physical points on the Earth's surface. The main physical and mathematical properties of a \ac{TRS} (at the definition and conventions level) or of the \ac{TRF} (at the realization level) include each its origin, scale, orientation, and their time evolution. The \ac{CM} of the Earth System, or geocenter, as the realized origin of the \ac{TRF} on long-term scales, needs to be accurately determined including its temporal motion \citep[e.g.,][]{petit2010}. The temporal variations of the geocenter represent a component of mass change (at spherical harmonic degree one) that is not directly observable from a mass-change mission such as GRACE-FO \citep{wu_geocenter_2012}. While the degree one component of mass change can be derived from a combination of GRACE data with ocean model output \citep[e.g.,][]{swenson2008a,sun_optimizing_2016, sun_statistically_2017} or space geodetic techniques such as \ac{GNSS}, \ac{SLR} \citep[e.g.,][]{fritsche2009a, glaser2015a}, a high-quality \ac{TRF} solution furnished by space geodesy that allows a matching with the temporal resolution of the GRACE-FO data would be highly desired (see Sect.~\ref{sec:longwave} for more details).

Any bias or drift in the \ac{TRF} components propagates into the estimated parameters based on the reference frame. It concerns for instance the measurements of vertical land motion and crustal deformation. This encompasses geological hazards but also human-induced effects (subsidence of land and coastal areas due to different effects) or even measurement of ongoing coastal erosion \citep{national_academies_of_sciences_evolving_2020}. Further examples are \ac{GIA} or mean sea level variability in space and time \citep{king2010a,collilieux2011a}. The global sea level rise of about \SI{3.7}{mm.year^{-1}} \citep{ipcc2021} is numerically small, but well ascertained within the range of measurement accuracy. In addition, an acceleration in the rate of sea level rise has also been observed \citep[e.\,g.\,][]{nerem2018a,veng2021a}. 
In areas at risk of flooding, the global mean sea level rise is a main topic of public dispute and political decisions. For its accurate monitoring and reliable prediction, the accuracy and long-term stability of the \ac{TRF} should be at least one order of magnitude better than the observed effects, leading to the GGOS accuracy and stability goals.

As stated in the introduction, the importance of accurate and stable reference frames was highlighted by the \ac{UN} \citep{UN2015a}. The implementation of the \ac{GGRF} is intended to support the increasing demand for positioning, navigation, timing, mapping, and other geoscientific applications. Indeed, the \ac{GGRF} is essential for a reliable determination of changes in the Earth system, for natural disaster management, for monitoring sea level rise and climate change, and to provide accurate information for decision makers. Furthermore, due to globalization and interoperability requirements, there is a growing demand for spatial data infrastructure. Precise spatial information is needed in many areas beneficial to society, including transportation, construction, infrastructure, process control, surveying and mapping, and Earth sciences, and is especially important for monitoring progress towards the \ac{UN}' \ac{SDGs} \citep{url-SDG} \citep[see, e.g.][]{UN-GGIM-SDGs}.

\subsection{Present status of the terrestrial reference frame realization} \label{sec:ITRSrealization} %

The actual realization of the \ac{ITRS}, accessible to the users, is the \ac{ITRF}. The computation of the \ac{ITRF} is based on a rigorous combination of different \ac{TRF} solutions provided by the four space geodetic techniques (\ac{DORIS}, \ac{GNSS}, \ac{SLR}, \ac{VLBI}), as well as the terrestrial local tie measurements conducted at co-location sites where two or more geodetic instruments operate. Local ties are the relative coordinates between the reference points of the individual instruments. They are crucial to connect the exact points of the observations of the different techniques in the \ac{ITRF} construction. The \ac{ITRF} is provided to the users in the form of station positions at a reference epoch and corresponding linear station velocities, and since the ITRF2014, parametric models for sites subject to major earthquakes \citep{altamimi2016}. The ITRF2020 was published in April, 2022 \citep{url-ITRF}. It now provides seasonal signals caused mainly by loading effects, expressed in both the Earth's \ac{CM} frame as sensed by \ac{SLR}, but also in the \ac{CF} frame \citep{altamimi_itrf2020_2021, Altamimi_2022}.

The \ac{ITRF} long-term origin is defined by \ac{SLR}, the most accurate satellite technique in sensing the Earth's \ac{CM}. The \ac{ITRF} long-term scale, however, is defined by an average of the \ac{SLR} and \ac{VLBI} intrinsic scales. The consistency of these scales still needs to be improved, since both techniques are subject to systematic errors and other technical limitations, such as time and range biases for \ac{SLR}, antenna deformation for \ac{VLBI}, etc. The GENESIS mission will help to solve these inconsistencies. The \ac{ITRF} orientation and its time evolution are defined to be the same for the successive \ac{ITRF} realizations.

Although the \ac{ITRF} is the most accurate \ac{TRF} available today, it still needs at least an order of magnitude of improvement in order to meet the scientific challenges of observing Earth system variability. The \ac{ITRF} is not only a fundamental standard for Earth science applications, but its elaboration, using extensive data analysis, also allows to evaluate the level of consistency between space geodetic techniques and to assess the systematic differences that are a major limiting factor in the \ac{ITRF} accuracy.

The analysis of the input data submitted to the latest \ac{ITRF} version, ITRF2020 \citep{itrf_itrf2020_2022}, was the occasion to re-evaluate the current level of consistency among the four main space geodetic techniques and their strengths and weaknesses. There are still a number of factors that limit the \ac{ITRF} accuracy, as shown (or reconfirmed) by the ITRF2020 results.

Although the ITRF2020 long-term origin is defined solely by \ac{SLR}, weaknesses in its realization include the poor number and geometry of \ac{SLR} stations in operation today: the number of the most prolific \ac{SLR} stations does not exceed 16, and not all of these stations have the same level of performance. The only internal evaluation that can be made is the level of agreement between ITRF2020 and previous \ac{ITRF} versions, namely ITRF2005, ITRF2008, and ITRF2014 whose origins were also defined using SLR data submitted in the form of time series. ITRF2020 results indicate that the agreement in the origin components, with respect to the past three ITRF versions, is at the level of \SI{5}{mm} in offset and \SI{0.5}{mm/year} in rate, values that are still far away from the science requirements.



Counting the number of ITRF2020 co-locations between \ac{VLBI}, \ac{SLR} and \ac{DORIS}, 11 \ac{VLBI}--\ac{SLR}, 12 \ac{VLBI}--\ac{DORIS} and 11 \ac{SLR}--\ac{DORIS} ties exist. These numbers of co-locations are  too small to provide a reliable combination of these three techniques alone. The \ac{GNSS} network is fundamental in determining the \ac{ITRF} by connecting the three other techniques to \ac{GNSS}, since almost all \ac{SLR} and \ac{VLBI} stations, and about two thirds of the \ac{DORIS} stations are co-located with \ac{GNSS}. Only \SI{32}{\%} to \SI{50}{\%} of these co-location sites (with time spans $> 3$~years) have an agreement between the terrestrial tie vectors and the space geodetic estimates of better than \SI{5}{mm} in the three components. It is likely that most of the tie discrepancies (differences between terrestrial ties and space geodetic estimates) are caused by systematic errors in the techniques. The poor spatial distribution of co-location sites is also a major limiting factor for the present accuracy of the terrestrial reference frame realization, which would be tackled by the GENESIS mission.

The \ac{ITRS} Center of the \ac{IERS}, hosted by \ac{IGN} France, is responsible for the maintenance of the \ac{ITRS}/\ac{ITRF} and the official \ac{ITRF} solutions. Two other \ac{ITRS} combination centers are also generating combined solutions: \acl{DGFI-TUM} \citep[DGFI-TUM; ][]{seitz_2008_2012, seitz_dtrf2014_2022} and \acl{JPL} \citep[JPL; ][]{wu_kalrefkalman_2015, abbondanza_jtrf2014_2017}. These ITRS realizations provide a valuable possibility to validate official ITRF solutions and thus help to increase the reliability of the ITRF.

GENESIS, as a fully calibrated satellite-based platform, will provide a complementary co-location of the four techniques in space, a ``Core co-location site in space'', as an optimal supplement to the existing co-locations on ground. This is essential to identify and potentially reduce the systematic errors and/or determine whether the errors come from the terrestrial ties or from the space geodetic estimates. 

The GENESIS mission will improve our ability to simultaneously identify the systematic errors and to consequently improve the \ac{ITRF} accuracy and stability, particularly the origin and the scale that are the most critical parameters for scientific applications. GENESIS will leverage the crucial existing ground-based co-location network, allowing the development of future-proof terrestrial reference frames.

\subsection{Improvements in the ITRF geocenter and scale} \label{sec:GeocScale}%

We define herein the geocenter motion as the motion of the \acf{CM} of the whole Earth (solid body and fluid envelope) with respect to the geometrical \acf{CF} of its deformable terrestrial crust. This motion is strongest at the annual frequency (\SI{2}{mm} to \SI{3}{mm} in the equatorial plane and up to \SI{5}{mm} in the direction of the polar axis) where it mostly reflects non-tidal fluid mass redistribution on the Earth's surface. The long-term secular variation represented by a linear rate is believed to be less than \SI{1}{mm/year} \citep{metivier2010, metivier2011}. In addition, atmospheric, hydrologic and oceanic masses cause deformations of the Earth's surface due to loading effects \citep{wu_geocenter_2012}. The space geodetic stations tied to the crust, thus, show variations of their position (mainly in the height component) due to variations of the loading effects caused by major water and atmosphere mass transport occurring over large regions. These position variations lead to changes in the \ac{CM} with respect to the \ac{CF}. Determining the variations of the geocenter is therefore important for understanding the long-wavelength changes in the distribution of mass within the Earth's system (see Sect.~\ref{sec:longwave}).

\begin{figure}
\centering
\includegraphics[width=1.0\linewidth]{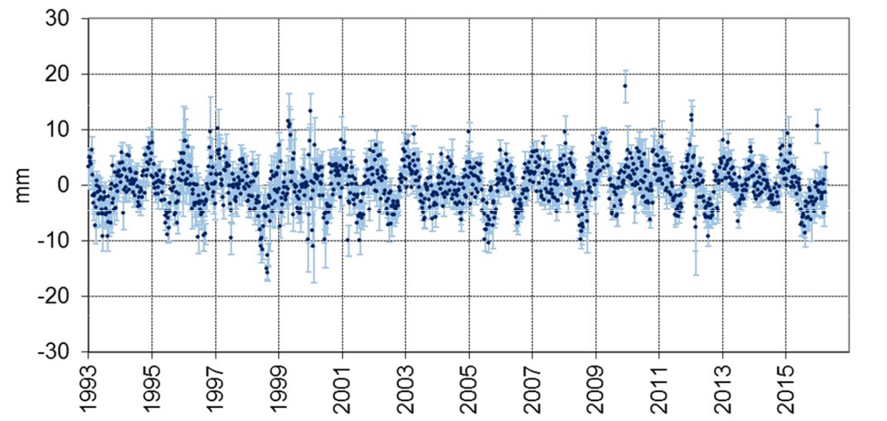}
\caption{\label{fig:SLR} \ac{SLR} derived geocenter motion, X component (ILRSA combined solution, ASI/CGS).}
\end{figure}

The geocenter motion is accessible by ground station observations (tied to the crust's \ac{CF}), used to observe the natural orbital motion of the satellites about the Earth's \ac{CM}. Yet, space geodetic observation of the geocenter motion is still in its infancy. Independent solutions derived using different techniques have systematic differences as large as the signal level. Estimating geocenter coordinates is one of the most demanding applications of high precision geodetic techniques due to the current precision of the geodetic data, and the nature and magnitude of different types of systematic error.

Up to now the geocenter motion is traditionally measured by \ac{SLR} using the observations to geodetic satellites (see Figure~\ref{fig:SLR}). The geodetic satellites such as LAGEOS or LARES are considered to be well suited for determining the geocenter motion owing to their mission characteristics, such as orbit altitude, low area-to-mass ratio, and thus minimized non-gravitational orbit perturbing forces. Until now, determination of geocenter coordinates based on the \ac{SLR} observations to active \ac{LEO} satellites was limited because of the issues in non-gravitational force modeling acting on \ac{LEO}s. In principle, the geocenter coordinates should be well determined from any satellite mission that is continuously observed and has dynamical orbits of superior quality. Therefore, GENESIS can introduce an alternative for the geocenter recovery w.r.t. passive geodetic satellites. 

However, the accuracy of \ac{SLR} data is extremely sensitive to the presence of observational biases therein, e.g., range biases and network effects \citep{collilieux_effect_2009}, affecting also the geocenter determination. Determining these biases would necessitate the use of an independent geodetic technique. Range biases are calculated in \ac{SLR} processing as additive constants in the modeled range, which should be in essence independent of the epoch of observation and measurement conditions, such as station elevation/azimuth angles, or measured range. However, the range biases not only compensate for ranging machine errors 
but also absorb the modeling errors such as satellite center of mass offsets, orbit force model deficiencies, or tropospheric delay \citep{appleby_assessment_2016, luceri_systematic_2019, drozdzewski_tropospheric_2021}. The presence of ambiguous range biases corrupts the estimation of fundamental geodetic products including site coordinates, the terrestrial reference frame scale and origin (geocenter motion), and geocentric gravitational constant (GM) \citep{couhert2020b}. GENESIS can expand our knowledge on the range biases aiming to improve the consistency of \ac{SLR} geodetic products with the other space and satellite geodetic techniques.

\ac{GNSS}-based determinations of the geocenter motion suffer from orbit modelling deficiencies due to an inherent coupling of the \ac{GNSS} orbit dynamic parameters: the \ac{GNSS} geocenter Z-component is strongly correlated with the parameterization of the \ac{SRP} \citep{meindl_geocenter_2013}. With only limited a priori knowledge about the non-conservative forces acting on \ac{GNSS} satellites, we must incorporate additional empirical orbit parameters into the solution, i.e., Empirical CODE Orbit Model or Jet Propulsion Laboratory GSPM. The errors in the orbit model, as well as the correlations between the estimated parameters \citep{rebischung_collinearity_2014}, introduce spurious orbit-related signals in the \ac{GNSS}-based geocenter motion estimates \citep{meindl_geocenter_2013, rodriguez-solano_reducing_2014}. The consistency between \ac{GNSS}-based and \ac{SLR}-based geocenter motion estimates can be improved by using satellite macromodels \citep{zajdel_geocenter_2021}. Another way to improve the \ac{GNSS}-based geocenter motion is the combined multi-\ac{GNSS} processing \citep{scaramuzza_dependency_2018} or the inclusion of Galileo satellites on an eccentric plane \citep{zajdel_geocenter_2021}.




Other approaches \citep{haines_realizing_2015, mannel_geocenter_2017, kuang2019, couhert2020} demonstrated the possibility to observe the geocenter motion with \ac{GNSS} tracking data in addition to \ac{LEO} satellites (e.g., GRACE, GOCE, or Jason-like satellites), which helped to reduce the errors coming from the \ac{GNSS}-only determination. Such methods would be well suited to derive \ac{GNSS}-based geocenter time series with GENESIS. The synchronization of the GENESIS onboard \ac{USO} thanks to an \ac{A-LRR} would allow a modelling of the clock instead of estimating clock correction parameters for each epoch. This is expected to improve the capability for accessing the geocenter.

\ac{DORIS} as the third satellite technique is in principle also sensitive to the \ac{CM} of the Earth. \ac{DORIS} benefits from the well-distributed network of stations but trails other geodetic techniques in terms of the quality of station coordinates because of the limitation of non-gravitation perturbing forces modeling and precise orbit determination of active satellites equipped with \ac{DORIS} receivers. 
Moreover, the problems mentioned for \ac{GNSS} also apply to the \ac{DORIS} system. Yet, \ac{SRP} modelling error on the Jason-type satellites can be identified and mitigated without compromising the Z geocenter estimate \citep{couhert2018}.

\ac{VLBI} in its current application is a purely geometric technique, thus, it has no connection to the Earth's gravity field (including the \ac{CM} of the Earth). \ac{VLBI} can currently be connected to the satellite techniques only via the station network and the local ties, and is not able to contribute to the geocenter determination. However, numerical simulations demonstrated that geodetic \ac{VLBI} is able to observe geocenter motion using observations of Galileo satellites \citep{klopotek_geodetic_2020}, suggesting that the GENESIS mission will enable a VLBI-contribution to the estimation of geocenter motion.

For the current \ac{ITRS}, the origin is assumed to be aligned to the long-term Earth's \ac{CM}. In parallel, the geopotential models assume that on average the Earth's \ac{CM} is at the centre of the geodetic network (i.e., zero values for degree-1 geopotential coefficients). Thus, the importance of an accurate geocenter motion cannot be overstated. Not accounting properly for the geocenter motion affects both satellite altimetry, precise orbit determination and satellite-derived estimates of the change in regional mean sea level. Because of climate change, and the need to both measure the change in the ice-sheets and understand their impact on sea level and global fluid mass redistribution, we must explore strategies to better observe and model these subtle variations in the Earth's geocenter.

GENESIS is a unique opportunity to properly calibrate the space geodetic techniques against each other. By this, GENESIS helps to improve our understanding of the aforementioned systematic differences between geocenter solutions derived using independent techniques, allowing the best possible accuracy in the recovery of the geocenter time series. In addition, the \ac{VLBI} tracking of GENESIS is a unique opportunity to attach also the \ac{VLBI} technique to the \ac{CM} of the Earth. As a result, the \ac{TRF} determined from the GENESIS measurements will realize the origin located in the \ac{CM} of the Earth consistently for all four space geodetic techniques for the first time.

The same reasoning holds true for another fundamental property of the \ac{ITRF}, i.e., the scale that is currently determined by means of \ac{SLR} as well as by \ac{VLBI} data analysis. The scale is defined in such a way that there exists no scale factor and no scale factor rate with respect to the mean of \ac{VLBI} and \ac{SLR} long-term solutions as obtained by stacking their respective time series. But similar to the problems mentioned related to geocenter, \ac{SLR} and \ac{VLBI} suffer from systematics also affecting the scale, e.g., unknown range biases for \ac{SLR} stations and unknown antenna deformations for \ac{VLBI}. Due to the very limited number of co-located \ac{SLR}-\ac{VLBI} stations on the Earth's surface and due to the lack of a common satellite, the agreement of the \ac{SLR}- and \ac{VLBI}-derived scales is difficult to assess, and the two techniques cannot be well calibrated against each other. In ITRF2020 the scale discrepancy between \ac{SLR} and \ac{VLBI} could be significantly reduced, however only selected \ac{VLBI} sessions until 2013.75 and \ac{SLR} observations from 1997 until 2021.0 were used due to trends and jumps with unknown nature. GENESIS will probably help to reveal the reasons for them and improve the scale realization significantly.

Undisclosed information about the ground calibrations of the satellite antenna \ac{PCO} prevented the use of \ac{GNSS} for the scale determination. Thanks to the release of the Galileo phase center calibrations for both the ground (receivers) and space (satellite antennas) segment \citep{gsa_galileo_2017}, \ac{GNSS} became a new potential contributor to the realization of the terrestrial reference frame scale of the future \ac{ITRF} releases. \citet{villigerGNSSScaleDetermination2020} reported that the Galileo scale difference w.r.t. ITRF2014 is \SI{1.4}{ppb} at the epoch of 1st January 2018. The information about the satellite phase center calibrations, which have been published in 2019 by CSNO (China Satellite Navigation Office) for the BeiDou satellites, opened up a space for the second \ac{GNSS} able to provide an independent realization of the terrestrial reference frame scale \citep{zajdel_potential_2022}. However, some results indicate that the scales derived with BeiDou-released and Galileo-released satellite phase center calibrations are not consistent, and the bias between both reaches \SI{1.8}{ppb} \citep{qu_phase_2021}.

Studies have already demonstrated that the \ac{SLR}-based scale can be well transferred to the \ac{GNSS} network, if a satellite is used as co-location platform \citep{thaller_combination_2011, thaller_geocenter_2014}. In these studies, the \ac{GNSS} satellites tracked by \ac{SLR} were employed as co-location platforms, however, being limited to \ac{GNSS} and \ac{SLR} only. At the moment, \ac{DORIS} is unable to deliver reliable scale information due to uncalibrated or not well-calibrated antenna phase centre locations and variations (for ground stations as well as satellites). 

GENESIS will enlarge the satellite co-location to all four space geodetic techniques allowing cross-calibration of all techniques to determine a homogeneous scale. Thanks to the common platform on board GENESIS, the scale will be transferrable to all techniques, resulting in the best possible materialization of the \ac{ITRF}.

\subsection{Unification of reference frames and Earth rotation} %
\label{sec:UnifFrame}

Geodetic \ac{VLBI} uses the emission by extragalactic radio sources with well-defined positions in the sky. If it is possible to transmit a quasar-like signal from an orbiting platform with a precise orbit, then we would be able to better understand biases between the \ac{CRF} realized with positions of extragalactic radio sources and dynamical realizations by satellite orbits.

\ac{CRF}, \ac{TRF}, and the \ac{EOP} that describe the transformation between these two frames are fundamental for any kind of positioning on the Earth and in space and provide most valuable information about the Earth system. The \ac{ICRS} is a quasi-inertial reference system defined by extragalactic radio sources, mostly quasars, billions of light years away, and is realized as \ac{ICRF} with a set of quasar coordinates with a noise floor of about \SI{30}{\micro{}as} \citep{Charlot2020}. The positions of a set of globally distributed radio telescopes are determined using the difference in the arrival times of the signals at the different telescopes \citep{Sovers1998}.

The \ac{VLBI} technique provides direct access to the \ac{ICRS} and is the best technique for observing the full set of \ac{EOP}. Specifically, \ac{VLBI} is the only technique able to determine the position of the celestial intermediate pole in the \ac{ICRF}, expressed as celestial pole offsets to a conventional precession/nutation model, and the Earth's rotation angle, typically referred to as Universal Time or UT1-UTC. Table \ref{tab:unification} summarizes the parameter types and the space geodetic techniques contributing to their determination. The table also shows the parameters that can be used for a co-location of the techniques, both, on the surface of the Earth and in space. Satellite techniques rely on measurements between stations on the Earth's surface and satellites, whose orbits are subject to various gravitational and non-gravitational forces (e.g., \ac{SRP}). As a consequence, \ac{SLR}, \ac{GNSS} and \ac{DORIS} depend on a reference frame that is dynamically realized by satellite orbits and thus completely different in nature from the kinematic realization of the \ac{ICRS} by \ac{VLBI}. Presently, the only physical connection between the \ac{VLBI} frame and frames of \ac{SLR}, \ac{GNSS} and \ac{DORIS} is via the local ties on the ground; however, these ties reveal significant discrepancies with respect to the terrestrial frames delivered by the individual space geodetic techniques. 

GENESIS will link all the technique frames in space (see gray rows in Table \ref{tab:unification}). This concept and its realization will represent a breakthrough in improving the accuracy and consistency of the reference frame. In addition, GENESIS will also directly link the dynamical satellite frame to the quasars using differential \ac{VLBI} observations (D-VLBI), i.e. differencing the radio signal emitted by GENESIS with the signals from the fixed radio sources, the quasars.

\begin{table*}[htb!]
    \caption{Classification of estimated parameters derived from space geodetic techniques. Co-location in space with GENESIS is marked in gray (adopted from \citealp{maennel2016}).}
    \vspace{0.2 cm}
    \centering
    \begin{tabularx}{0.9\textwidth}{l|l|l|c|c|c|c|c} \hline
    Classification & Type & Parameter & \ac{VLBI} & \ac{GNSS} & \ac{SLR} & \ac{DORIS} & \ac{LLR} \\ \hline
    Common, & Satellite orbits & \ac{GNSS} orbits & ($\checkmark$) & $\checkmark$ & ($\checkmark$) & & \\
    global & & LEO orbit & & $\checkmark$ & $\checkmark$ & $\checkmark$ &  \\
    & & LEO clock corrections & & $\checkmark$ & & ($\checkmark$) &  \\
    & & \cellcolor{lightgray} GENESIS orbit & $\checkmark$ & $\checkmark$ & $\checkmark$ & $\checkmark$ &  \\
    & & \cellcolor{lightgray} GENESIS clock corr. & $\checkmark$ & $\checkmark$ & $\checkmark$ & $\checkmark$ &  \\
    & \ac{EOP} & Pole coordinates & $\checkmark$ & $\checkmark$ & $\checkmark$ & $\checkmark$ & ($\checkmark$) \\
    & & UT1 & $\checkmark$ & & & & ($\checkmark$) \\    
    & & LoD & $\checkmark$ & $\checkmark$ & $\checkmark$ & $\checkmark$ & ($\checkmark$) \\
    & & Nutation & $\checkmark$ & & & & ($\checkmark$) \\    
    & & Nutation rates & $\checkmark$ & $\checkmark$ & $\checkmark$ & $\checkmark$ & ($\checkmark$)   \\
    & Gravity field & Earth center of mass & & ($\checkmark$) & $\checkmark$ & ($\checkmark$) &  \\
    & & Low degree coefficients & & $\checkmark$ & $\checkmark$ &  $\checkmark$ & ($\checkmark$) \\   
    & TRF & Scale & $\checkmark$ & ($\checkmark$) & $\checkmark$ & ($\checkmark$) & ($\checkmark$)  \\ \hline
    Common, & Atmosphere & Ionosphere parameters & $\checkmark$ & $\checkmark$ & & $\checkmark$ & \\
    local & & Troposphere parameters & $\checkmark$ & $\checkmark$ & ($\checkmark$) & $\checkmark$ & \\
    & TRF & Station positions & $\checkmark$ & $\checkmark$ &  $\checkmark$ & $\checkmark$ & ($\checkmark$) \\
    & & Station velocities & $\checkmark$ & $\checkmark$ &  $\checkmark$ & $\checkmark$ & ($\checkmark$) \\
    & Time-Frequency & Station clock corrections & $\checkmark$ & $\checkmark$ &  ($\checkmark$) & $\checkmark$ & ($\checkmark$)  \\ \hline
    Technique- & CRF & Quasar positions & $\checkmark$ & & & & \\
    specific & & Moon orbit & & & & & $\checkmark$ \\
    & Instrumental & \ac{GNSS} clock corrections & & $\checkmark$ & & & \\
    & & Range biases & & & $\checkmark$ & & $\checkmark$ \\ \hline
    \end{tabularx}
    \label{tab:unification}
\end{table*}

Another issue of the \ac{ICRF} and \ac{ITRF} is that the realizations of the frames are independent of one another. The \ac{ITRF} is fixed when computing the \ac{ICRF} and vice versa. The approach leads to inconsistencies that map into the \ac{EOP} that connect the two frames. \ac{IUGG} Resolution No. 3 (2011) recommends that the highest consistency between the \ac{ICRF}, the \ac{ITRF} and the \ac{EOP} should be a primary goal in all future realizations of the \ac{ICRS}. Although the \ac{IUGG} recommendation has not yet been fulfilled, research in this direction has been initiated and simultaneous estimation of \ac{CRF}, \ac{TRF}, and \ac{EOP} have been achieved \citep{Seitz2014,kwak2018}. At the international level, this topic is being addressed by the ICRF3 Working Group of the International Astronomical Union and by Sub-Commission 1.4 on the Interaction of Celestial and Terrestrial Reference Frames of the \ac{IAG}. The common adjustment of the celestial and terrestrial reference frames and \ac{EOP} will strongly benefit from new observations provided by GENESIS.

In summary, to strengthen the link between \ac{VLBI} and the satellite techniques, it is imperative to use improved and better ties than what is currently available, i.e., the local ties at the relatively few \ac{VLBI}/\ac{GNSS} co-located sites. Initial work with dedicated space tie satellites demonstrated the feasibility of this approach, see simulations by \citet{anderson2018} or \citet{klopotek_geodetic_2020} and real observations to the APOD-A satellite by \citet{hellerschmied2018APOD}. Dedicated \ac{VLBI} beacons transmitting at \ac{VLBI} frequencies on a well-calibrated satellite such as GENESIS will enable the observation of satellites with \ac{VLBI} radio telescopes.

Moreover, GENESIS will not only combine \ac{GNSS} and \ac{VLBI}, but also \ac{SLR} and DORIS. Table \ref{tab:unification} illustrates that a rigorous combination of all the observation techniques and of as many of the common parameters as possible should be envisaged to overcome the weaknesses of the individual space geodetic techniques. GENESIS is a crucial element for improving the relationship between the reference frames of \ac{VLBI} and the satellite-based techniques.

The time variations of Earth rotation parameters contain subtle information about the mass transport in the system made up of the solid Earth, the external fluid layers, and the outer and inner core. An accurate determination of \ac{EOP} has long been at the origin of challenging studies related to the Earth's interior: e.g., insights into the coupling mechanisms at core-mantle and core-inner core boundaries by inversion of nutation data \citep{dehant2017} and to climate: e.g., mechanisms of angular momentum exchange between the solid Earth and the atmosphere-ocean system, link with climate change \citep{dickey2011}. In addition, the improvement of the \ac{VLBI} \ac{CRF} \citep{Charlot2020} contrasted with the optical data from the \ac{ESA}'s Gaia astrometry mission \citep{mignard2016} will allow to shed new lights on the physics of active galactic nuclei and quasars that will benefit from an improved stability of the radio frame currently limited to \SI{0.03}{mas}. 

Given that GENESIS will provide a direct link between the kinematic (\ac{VLBI}, quasar-based) and dynamic (satellite-based) reference frames and is expected, thus, to improve the consistency of the \ac{TRF}, \ac{CRF}, and \ac{EOP} realizations, all the above scientific domains will be positively impacted, extending thus the challenges of GENESIS well beyond its first scope.


The continuous tracking of the GENESIS satellite and the connection to \ac{GNSS} satellites will allow for mitigation of some technique-specific systematic effects currently observed in the GNSS-derived Earth rotation parameters. These errors come from constellation repeatability and orbital resonances between Earth rotation and satellite revolution period, as well as draconitic errors due to the limitations in the \ac{POD} of \ac{GNSS} satellites \citep{zajdel2020}. The integrated adjustment of low-orbiting GENESIS and \ac{GNSS} constellation could allow for the mitigation of technique-specific systematic effects observed in the polar motion and length-of-day variations, and thus, improve the quality of EOPs. 

The sub-daily Earth rotation can be monitored using space geodetic techniques. However, the current empirical sub-daily \ac{EOP} models derived from \ac{GNSS} or \ac{VLBI} differ from the geophysical models derived from ocean tides \citep{zajdel2021}. The sub-daily changes of the pole position are mainly caused by ocean tides, and to a smaller extent, by the atmosphere. However, GNSS cannot provide suitable values of some tidal constituents equal to half and one sidereal day due to the similar revolution period of the satellites. 

GENESIS, with its completely different orbit characteristics than the \ac{GNSS} satellites, will introduce an ever bigger step in this direction. Therefore, the sub-daily polar motion, libration terms, or sub-daily length-of-day variations will be better understood. Due to the continuous tracking of GENESIS, the derivation of sub-daily variations will be possible based on integrated observational techniques. Moreover, GENESIS will help in deriving sub-daily variations of the pole caused by the mass redistribution in the atmosphere, which are currently affected by large determination errors. Therefore, GENESIS will pave new opportunities for better understanding the sub-daily Earth rotation and relate their causes to the geophysical processes.  

\section{Benefits for Earth sciences}
\subsection{Long-wavelength gravity field} %
\label{sec:longwave}

Changes in the Earth's gravity field provide information about the redistribution of mass within the Earth system. These changes are measured with exquisite precision by dedicated gravity satellites such as GRACE-FO. Historically, geodetic tracking data (primarily \ac{SLR} data, but also \ac{DORIS} and \ac{GNSS}) has also been used to provide information about the long-wavelength (low-degree) time-variable gravity field \citep[e.g.,][]{cheng_grace_2013, cerri_doris-based_2013, richter_reconstructing_2021, sosnica2015,blossfeld2018}. Due to issues with the accelerometers and possibly with tidal aliasing the C20 and C30 solutions from GRACE and GRACE-FO so far have been supplied by \ac{SLR} \citep{loomis_improved_2019, loomis_replacing_2020}. Because of these accelerometer issues, it remains important to monitor and inter-compare GRACE and GRACE-FO based solutions with independent solutions \citep[e.g.,][]{chen_assessment_2021}, where that is possible at the longest wavelengths.

GENESIS can provide independent estimates of the low-degree Stokes coefficients based on all geodetic techniques. \ac{SLR} studies and simulations showed that adding one satellite to a solution based on five \ac{SLR} satellites may significantly improve the determination of the low-degree spherical harmonics of the Earth gravity field \citep{blossfeld2018,kehm2018}. The main improvements were seen in C10, C20, and C40, the standard deviations of which are improved up to \SI{30}{\%}. Also, observations of GENESIS with \ac{VLBI} would strengthen the integration of the Earth geometry, rotation and gravitational field. The Stokes coefficients are common parameters to all techniques such as a subgroup of the \ac{EOP}, namely the terrestrial pole coordinates and its first derivatives (see Table~\ref{tab:unification}).

Moreover, GENESIS will be of benefit in the refinement of the Earth GM, which helps to define the scale of the \ac{TRF}. The current value of \SI{3.986004415e14}{m^3.s^{-2}} was determined by \citet{ries_progress_1992} and has an uncertainty of the order \SI{2}{ppb}, corresponding to \SI{\pm 2}{cm} in the absolute radial position of high-orbiting \ac{GNSS} satellites. Some recent work has suggested possible solutions closer to \SI{3.986004418e14}{m^3.s^{-2}} \citep{couhert2020b}. 

An exciting prospect is that the precise tracking on GENESIS by multiple geodetic techniques, together with \ac{SLR} tracking to the satellites LARES and LARES-2 could lead to improve Earth GM determination with a much lower uncertainty. 

\subsection{Altimetry and sea level rise} %
\label{sec:AltiSea}

\begin{figure}
\centering
\includegraphics[width=1.0\linewidth]{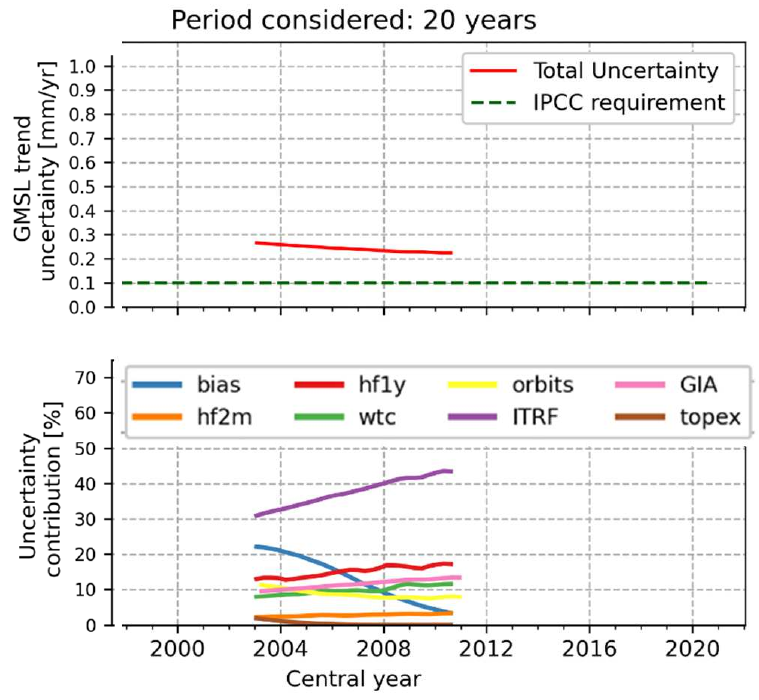}
\caption{\label{fig:GMSLtrend} Total uncertainty in the \acf{GMSL} trend and individual contributions  \citep{blazquez_importance_2022}. The uncertainty coming from the \acf{ITRF} is clearly the major contribution, which implies a total uncertainty currently well over the \acf{IPCC} requirement. Updated from \cite{ablain_uncertainty_2019}}
\end{figure}

Measurements of sea level rise over the last century (and more) have been derived from observations of sea level change as recorded by a global distribution of tide gauges. Tide gauge observations are relative observations, i.e., the changes in sea level are provided relative to the land to which they are attached. But processes such as glacial isostatic adjustment and tectonics cause the land to move in the vertical direction. To measure sea level change in a global reference frame, the tide gauges are geodetically tied to the \ac{TRF} using co-located \ac{GNSS} stations. Thus, the ability to observe contemporary sea level rise at global or local scales is limited by the stability of the terrestrial reference frame, i.e., the metrological basis for the determination of the vertical motions of stations on the Earth’s surface. The reference frame stability is one of the major error sources in the determination of global and regional sea level rise \citep{beckley_reassessment_2007, blewitt_geodetic_2010, blazquez_importance_2022}, as illustrated in Figure~\ref{fig:GMSLtrend}.

As recognized by \ac{GGOS}, sea level poses the most stringent requirements on the accuracy and stability of the \ac{TRF}. In addition to the tide gauge problem outlined above, precise satellite orbits in a highly stable and accurate \ac{TRF} are crucial to observe sea level using satellite radar altimetry. Since the launch of the TOPEX/Poseidon satellite in 1992, followed later by the Jason and Sentinel series of altimetry satellites, sea level variations have been routinely observed from space while tide gauges are also required to detect drifts in satellite radar altimetry data \citep{abdalla_altimetry_2021}. Moreover, it is acknowledged that the continuous improvement of  satellite \ac{POD}, through the precision and quality of the tracking systems, reference frame, Earth Rotation Parameters, and static and time-variable geopotential models are crucial in order to reach the specifications of altimetry missions \citep{abdalla_altimetry_2021}.


To be useful in long-term sea level studies, sea level and the vertical land motion should be measured in a reference frame at least one order of magnitude more accurate than the contemporary climate change signals of \SI{1}{mm/year} to \SI{3}{mm/year} observed on average in sea level records, either from tide gauges or satellite radar altimetry, leading to the GGOS accuracy and stability goals. The GGOS stability goal of \SI{0.1}{mm/year} of global and regional sea level variations over several decades can only be achieved by more accurate and more stable reference frame realizations, which is the primary goal of the GENESIS mission.

\subsection{Determination of ice mass loss} %
\label{sec:IceMass}

Direct local observations and space geodetic techniques including gravimetry, radar and laser altimetry, optical and synthetic aperture radar imagery and \ac{GNSS}, have provided clear evidence for large changes in the world's glaciers and ice sheets, in response to present climate change \citep[e.g.,][in press]{shepherd_mass_2018,shepherd_mass_2020,millan2022,foxkemperIPCC}. However, despite the extensive literature on the subject, the ice mass balances over the different ice sheets and smaller glacier regions are associated with large uncertainties  \citep[e.g.,][]{cazenave_global_2018, metivier2010,khan2015}. In particular, the question of possible local accelerations of ice mass loss in Greenland is still open \citep[e.g.,][]{velicogna2013, velicogna2014, velicogna2020}.

Direct observations of glaciers and ice sheets are local and only partially resolved in time, while space observations provide insights into the cryosphere evolution at global scale and at regular timescales. Space altimetry \citep[e.g., ICESat, ICESat-2, CryoSat-2 missions; e.g.,][]{felikson2017,sorensen2018} gives high resolution observations of the ice surface elevation, with a relatively poor time resolution. 
These techniques, unfortunately, cannot provide directly ice mass balances because the mean density of the ice column is not known and may largely vary locally due to compaction processes within the firn layers of ice sheets \citep{medley2020}. Additional information is therefore mandatory, generally based on compaction assumptions and climate models \citep{huss2013,kuipers2015,medley2020}.
Space gravimetry from GRACE and GRACE-FO missions provides direct information on mass variations, on a monthly basis, but with a spatial resolution of a few hundred kilometers \citep{tapley_grace_2004,landerer_extending_2020}. However, the various contributions from the solid Earth and the surface layers cannot be separated with space gravimetry data alone.
These other signals have to be removed before glacier and ice sheet signals can be retrieved, specially in the regions where these signals are of the same order of magnitude or even higher \citep{wouters_global_2019}. Removing these signals implies adding new information from models as for the \ac{GIA} or observations and adding new sources of uncertainty. 
In particular, the separation between the recent ice melting signals and the \ac{GIA} induced by the last glacial period \citep[e.g.,][]{whitehouse_glacial_2021,peltier2015,lambeck2014}, or by the little ice age \citep{kjeldsen2015}, is also a complex issue. However, \ac{GIA} modelling approaches today depend also on space geodetic observations \citep[e.g.,][]{argus2021,khan2016} and therefore would benefit from a better estimation of \ac{ITRF} parameters.
Of particular interest is the uncertainty associated with the \ac{GIA} in Antarctica, responsible of \SI{20}{\%} of the uncertainty in the Antarctica mass change for the period 2005-2015 \citep{blazquez_exploring_2018}.

Space geodetic techniques, in particular \ac{GNSS}, provide also useful information on ice sheets and glaciers evolution, by showing the ground deformation at geodetic stations induced by local ice mass changes \citep[e.g.][]{khan2016,whitehouse2019}. However, \ac{GNSS}-based velocities are sensitive to very local changes in the ice sheets, which make it difficult to confront with more global approaches such as space altimetry and gravimetry \citep[e.g.][]{khan2010}.
Other techniques are also promising for monitoring ice caps evolutions, such as radar interferometry. A combination of all techniques is today the best way for monitoring ice sheet changes, however discrepancies are clearly evidenced \citep{shepherd_mass_2018,shepherd_mass_2020}.

\begin{table}[hb]
    \caption{Uncertainty in the water mass change induced by an uncertainty of 1 mm in each axis of the geocenter motion.}
    \centering
    \begin{tabular}{l|r r r}
         &\multicolumn{1}{c}{Ocean Mass}&\multicolumn{1}{c}{Greenland}&\multicolumn{1}{c}{Antarctica}  \\
        & mm SLE& Gt&Gt\\
        \hline\\
    X& 0.5 & $<$1 & $<1$\\
    Y& 0.3 & $<$1&7\\
    Z& 0.6 &11 & 68\\
    \hline\\
    \multicolumn{4}{r}{Updated from \cite{blazquez_exploring_2018}}\\
    \end{tabular}
    \label{tab:geocenter_effet}
\end{table}

All space geodetic techniques rely on the availability of a precise and stable terrestrial reference frame such as ITRF2014 and ITRF2020 \citep[see Sect.~\ref{sec:ITRSrealization};][]{altamimi2016}. The stability over time of such a frame may inevitably impact all kind of geophysical interpretations that are deduced from geodetic observations. It was shown that it is not possible to ensure consistency between the \ac{ITRF}2008 origin and the mean \ac{CM} at a level better than \SI{0.5}{mm/year} \citep{wu2011}. These inconsistencies remain in ITRF2014 and cannot be properly explained by geophysical models \citep{riddell_uncertainty_2017}. Moreover, \ac{CM} motions are today more than ever difficult to estimate with precision and stability, because they are also impacted by climate change. It has been shown that the global ice sheet melting may induce today an accelerated \ac{CM} motion, possibly up to \SI{\sim 1}{mm/year} with respect to the \ac{CF}, towards south pole along the Earth's rotational axis \citep[e.g.,][]{metivier2010,metivier2011,metivier2020}. 

An error of a few tenths of \si{mm/year} in the frame origin stability estimation is well known to have a large impact on the orbit calculations of satellites and in the water mass redistribution on the surface (See Table \ref{tab:geocenter_effet}). Nowadays uncertainty in the long term trends in the geocenter motion of \SI{\pm 0.3}{mm/year} leads to uncertainties in the Antarctica mass change of \SI{18}{Gt/year} \citep{wu_geocenter_2012,blazquez_exploring_2018}.

As mentioned before, we expect that GENESIS will improve the determination of the reference frame. Such a stable \ac{ITRF} should drastically reduce the frame dependency of ice mass balance estimations. 
 
\subsection{Geodynamics, geophysics, natural hazards} %
\label{sec:geo}

Post-Glacial Rebound, also known as \ac{GIA}, is the delayed viscoelastic response of the solid Earth to unloading caused by the melting of ice after the last glacial maximum (\SI{\sim 13000}{years} ago). The induced vertical uplift can reach more than \SI{1}{cm/year} in North America and Fennoscandia, where the ice thickness was the largest. This long-term deformation depends on the ice coverage (both extent and thickness), the melt history, and the viscoelastic properties of the Earth's mantle. \ac{GIA} models currently are built using various observations, from moraines to relative sea level variations, Earth's oblateness variations, length-of-day variations, etc. Vertical rebound velocities from \ac{GNSS} permanent stations provide an independent but also absolute observation of the \ac{GIA} in high latitudes (where there is no ice today), and improve our knowledge of the Earth’s rheology at long timescales. Improving \ac{GIA}  models (Earth's rheology and ice history) requires better estimates of uplift rates which can only be achieved with a more precise and stable reference frame \citep[see, e.g.,][]{metivier_past_2020}.

On shorter timescales, \ac{GNSS} stations also record Earth's elastic response to surface mass redistribution within the climatic system (mainly continental water storage, atmosphere and ocean). Dense networks of permanent \ac{GNSS} stations can now be used to derive soil and snow water content at seasonal timescales, but has also provided evidence for extreme droughts, especially in California \citep[see, e.g.,][]{argus_seasonal_2014, fu_gps_2015, jiang_hydrological_2022}. \ac{GNSS} time series from dense networks can be used to refine the information provided by space gravimetry missions (GRACE and GRACE-FO) at longer spatial wavelengths (see Sect.~\ref{sec:longwave}). Amplitude and spatial extent of surface water mass variations can be inferred from both vertical and horizontal deformation measurements. In particular, horizontal displacements help to refine the determination of the location and the spatial extent of the load. This elastic Earth's response to surface loads has to be separated from a longer-term deformation, which can only be obtained with a more accurate and stable reference frame as proposed by the GENESIS project.

Observed ground movements at the Earth surface are manifold and related to a whole set of processes. Common and essential to all these movements are detection and monitoring to execute and develop risk assessment strategies. Natural hazards, such as earthquakes, volcanic hazards or landslides may be preceded by small displacements of the Earth’s surface. Dense networks of \ac{GNSS} stations in Japan, the western United States, and South America have been installed to monitor these surface displacements, related to the seismic cycle. In particular, pre-earthquake surface deformation can be related to the stress and the state of stress in the lithosphere. Surface displacements from increasing stress in the lithosphere may have small amplitudes. Therefore, a very stable and precise reference frame is required to be able to interpret these observations as reliable prediction tools for the onset of hazards versus errors in the techniques themselves.

\subsection{Top of atmosphere radiation budget and Earth energy imbalance}
\label{sec:RadBudget}

The radiative imbalance at the \ac{TOA} is the most fundamental metric to estimate the status of climate change. At equilibrium, the climate system receives as much visible energy from the sun as it emits infrared radiation towards space. Over the last decades, greenhouse gases and aerosol concentrations have been increasing in the atmosphere, blocking longwave radiation and leading to an imbalance at \ac{TOA} between the incoming solar radiation and the outgoing longwave radiation \citep{hansen_earths_2011, trenberth_challenges_2014}. This imbalance, known as the \ac{EEI}, is about \SI{0.5}{W.m^{-2}} to \SI{1}{W.m^{-2}} \citep[e.\,g.\,][]{loeb_clouds_2018}. It characterizes the general heat uptake of the climate system that is responsible for current climate change. It is particularly challenging to estimate the \ac{EEI} from \ac{TOA} radiation fluxes since it is 2 orders of magnitude smaller than the mean incoming solar radiation and the mean outgoing longwave radiation (\SI{\sim 340}{W.m^{-2}}) \citep{lecuyer_observed_2015}. The \ac{CERES} project has been measuring the Earth radiative budget at \ac{TOA} for several decades now \citep{loeb_clouds_2018}. The measurements are difficult, involving the incoming solar radiation, the scanning of outgoing radiation both visible and infrared, cloud cover, aerosols, and instrumental problems. The precision of the measurement is evaluated at the order of \SI{0.17}{W.m^{-2}} (\SI{90}{\%} confidence level) at interannual time scales but because of a potential bias of about \SI{\pm 2}{W.m^{-2}} the accuracy is above \SI{\pm 2}{W.m^{-2}}. The precision of \ac{CERES} is sufficient to evaluate small changes in time of the \ac{EEI} that are induced by natural or anthropogenic forcing \citep{loeb_ceres_2020,raghuraman_anthropogenic_2021}. But the accuracy is not sufficient to estimate the mean \ac{EEI} generated over the past decades by anthropogenic greenhouse gases emissions.

Another approach to estimate the \ac{EEI} consists in estimating the excess of energy that is stored in the climate system in response to the \ac{TOA} radiative imbalance. With its high thermal inertia and its large volume, the ocean accumulates, in the form of heat, more than \SI{90}{\%} of the excess of energy that is stored by the climate system \citep{von_schuckmann_heat_2020}. The other climate reservoirs (i.e. atmosphere, land, and cryosphere) play a minor role in the energy storage at seasonal and longer timescales \citep{von_schuckmann_heat_2020}. As a result, the \ac{OHU} is a precise proxy of the \ac{EEI} and estimating the \ac{OHU} is an efficient approach to estimate the \ac{EEI}. 

The \ac{OHU} can be estimated with an accuracy of a few tenths of \si{W.m^{-2}} and thus provides an approach to estimate the mean \ac{EEI} generated over the past decades by anthropogenic greenhouse gases emissions. This is possible with 2 approaches: (1) from direct in-situ measurements of temperature–salinity profiles mainly derived from the Argo float network; (2) from the thermal expansion of the ocean derived from a space geodetic approach \citep{meyssignac_measuring_2019}. These methods are complementary, with their own advantages and limitations. The direct measurement approach relies on in situ measurements from Argo which are unevenly spatially distributed with poor sampling of the deep ocean (below \SI{2000}{\meter} depth), marginal seas, and below seasonal sea ice. The space geodetic approach measures the sea level changes due to the thermal expansion and saline contraction of the ocean (also called sterodynamic sea level changes) derived from the differences between the total sea level change derived from satellite altimetry measurements and the barystatic sea level changes from satellite gravity measurements. This approach offers consistent spatial and temporal sampling of the ocean through time, with a nearly global coverage of the oceans, except for the polar regions (above \ang{82}). It also provides \ac{OHU} estimates from the entire ocean water column. But it does not provide the vertical structure of the \ac{OHU} unlike the Argo approach. It is crucial to develop both the geodetic approach and the in-situ approach to derive \ac{EEI} estimates that are cross validated and thus reliable.

The \ac{EEI} shows time variations in response to anthropogenic emissions and natural variability like ocean–atmosphere interactions or volcanic eruptions. The coupled natural variability of the ocean and of the atmosphere leads to monthly to interannual variations of the order of a few \si{W.m^{-2}} \citep{loeb_clouds_2018}. Decadal and longer-term variations of the order of a few tenths of \si{W.m^{-2}} are associated with the anthropogenic and the natural forcing of the climate system \citep{loeb_ceres_2020}. On decadal time scales the \ac{EEI} shows trends of the order of a few cents of \si{W.m^{-2}.year^{-1}} in response to changes in the external forcing  either natural or anthropogenic \citep{loeb_ceres_2020, raghuraman_anthropogenic_2021}. To evaluate these variations and particularly the small decadal and longer-term response of \ac{EEI} to anthropogenic or natural forcing, \ac{EEI} should be estimated with an accuracy better than \SI{\pm 0.1}{W.m^{-2}} and a stability better than \SI{\pm 0.02}{W.m^{-2}/year}. This is particularly challenging, and it requires a fine characterisation of the errors associated with the \ac{EEI} estimates. In the case of the \ac{EEI} derived from the geodetic approach the limiting factors at decadal time scales come from the uncertainty in the \ac{GIA}  correction and in the \ac{ITRF} realization \citep[see][]{blazquez_exploring_2018, meyssignac_measuring_2019, guerou_uncertainties_2022}. In particular, the uncertainty on the Z motion of the geocenter of the \ac{ITRF}  which affects both the satellite altimetry estimate of the sea level changes at mid to high latitudes and the gravimetry estimate of the ocean mass changes is a primary source of uncertainty on the \ac{EEI} at decadal and longer time scales \citep[see][]{marti_monitoring_2022, blazquez_exploring_2018, guerou_uncertainties_2022}.

To reach an accuracy of \SI{\pm 0.1}{W.m^{-2}} and a stability of \SI{\pm 0.02}{W.m^{-2}/year} in \ac{EEI} on decadal time scales an accuracy of \SI{\pm 0.25}{mm} and a stability of \SI{\pm 0.05}{mm/year} is necessary on sea level and ocean mass rates estimates at decadal time scales. This is achievable only with more accurate and more stable reference frame realizations, which is the primary goal of the GENESIS mission.

\subsection{Ionospheric and plasmaspheric density} \label{sec:IonoPlasma} %

The Earth's ionosphere is defined as the atmospheric layer, typically between \SI{80}{km} and \SI{1000}{km} altitude, where the electron density is sufficient to significantly influence the propagation of electromagnetic waves that travel into it \citep{davies1990}. Its main effect is the modification of the wave propagation velocity, which is proportional to the integrated electron density along the wave propagation direction, called \ac{TEC}. Secondary effects are a modification of the wave amplitude and a bending of the propagation vector, with respect to the generally assumed straight line. 

The electron density profile is characterized by a Chapman-profile shape, with a peak at an altitude typically ranging from \SI{200}{km} to \SI{350}{km}. At mid and low latitudes, above the ionosphere is the plasmasphere, constituting a reservoir of cold particles (mainly electrons, protons and helium ions) that fills during the day and drains at night \citep{russel2016}. Its location, as well as the particle motion, is controlled by the geomagnetic field whose dipolar nature confines the plasma. The main control of these ionized mediums is the solar activity that drives extreme ultraviolet radiation (hence the primary source for ionization) and that modifies the solar wind which constantly interacts with the geomagnetic field. Besides this variability ``from above''  many irregularities arise ``from below'', i.e. for which the origin is located in the lower atmospheric layers or at the ground level \citep{fullerrowell2017}. One of the most important ionospheric variability sources lies in the equatorial region and is known as the ``fountain effect'': a rapid change in the neutral wind creates an important $\vec E \times \vec B$ (where $\vec E$ is the electric field and $\vec B$ the magnetic filed) vertical drift that lifts the plasma up to \SI{1000}{km} altitude \citep{kelley2009}. The plasma plumes are then redistributed on the either side of the magnetic equator to form the so-called ``ionization anomaly crests'', being two regions of local maximum electron density. To that are associated small-scale irregularities called equatorial plasma bubbles, which appear during post-sunset hours at low-latitudes and produce plasma depletions that disturb radio communications and \ac{GNSS} services \citep{kintner2007}. More precisely, they are responsible for signal scattering that fades out the signal amplitude, leading to fluctuating signal-to-noise ratio that prevents optimal \ac{GNSS} satellite tracking, or even worst, interrupts it.

During the last two decades, an important number of \ac{GNSS} receivers has been included on board \ac{LEO} satellites orbiting at various altitudes, besides an extensive network of ground-based receivers. They continuously receive the signal broadcasted by \ac{GNSS} satellites offering an excellent time and space coverage and allowing to reliably monitor the \ac{TEC} above a given \ac{LEO} satellite \citep{Wautelet2017}. In the latter methodology, the retrieved \ac{TEC} is the by-product of the differential code biases computation performed using only \ac{LEO}-based observation (i.e. no ground station) to minimize the impact of the ionospheric peak in the differential code biases adjustment. Depending on the orbit altitude and the geomagnetic latitude, the \ac{TEC} above the spacecraft would mostly express the ionization crests or the plasmaspheric contribution. 

In the framework of a circular orbit at an altitude of \SI{6000}{km}, as planned for the GENESIS mission (see Sect.~\ref{sec:CDF}), the plasmaspheric TEC would be very small, and even negligible. The LEO-DCB (Differential Code Biases) computation software should be able to quantify this contribution and provide, if plasmaspheric TEC contribution is actually negligible, accurate and reliable DCB values for GNSS satellites and onboard receiver.



In addition to the zenith \ac{GNSS} antenna, a nadir-pointing \ac{GNSS} antenna would enable the observation of radio-occultation profiles of \ac{GNSS} satellites, as the same manner as for dedicated missions COSMIC and COSMIC-2. Using appropriate inversion methods on such observations will provide additional electron density profiles that will benefit the ionosphere/plasmasphere community. Moreover, the high inclination of the GENESIS orbit will provide occultations above polar and sub-polar regions, which are not geographically covered by the dedicated missions COSMIC and COSMIC-2. By providing electron density profiles at polar regions, GENESIS will improve the observability and the understanding of the dynamics in this region where the ionosphere meets the solar environment.

\section{Benefits for navigation sciences and metrology}
\subsection{Improvement in global positioning} %
\label{sec:GlobPos}

The positioning of stations in a \ac{GNSS} network relies on global solutions of complete satellite constellations and on the simultaneous adjustment of station coordinates. These solutions involve a large number of parameters that will degrade the observability of the station coordinates, and may also introduce biases in the solutions: tropospheric propagation models, empirical solar radiation pressure parameters of the satellites, antenna characteristics such as \ac{PCO}, \ac{PV}, which may or may not be adjusted.

In a global \ac{GNSS} solution, the clock biases of the transmitters and receivers must be managed. Regardless of the strategy used (estimation or differentiation), the resulting information for the geometry is the same. The result is a significant reduction in the observability of the Earth's center of mass, i.e. the origin of the reference frame \citep{rebischung_collinearity_2014, meindl_geocenter_2013}. This includes the motion of the geocenter whose north-south motion is not well observed by \ac{GNSS} constellations and is corrupted by draconitic signals from orbit solutions (see Sect.~\ref{sec:GeocScale}). 



Some improvements could be achieved if the GENESIS onboard oscillator can be modeled with sufficient accuracy over long periods (typically one day), allowing a drastic reduction in the impact of receiver-related clock parameters. A comparison of the onboard USO to ground atomic clocks could be achieved thanks to an \ac{A-LRR} (see Sec.~\ref{sec:laser}). For example, a model of the USO on board Jason-2 satellite was developed thanks to the \ac{T2L2} instrument \citep{belli_temperature_2016}.

The frame scale factor is also directly connected to the transmitters' and receivers' \ac{PCO}/\ac{PV} characteristics. For the ground receivers, independent calibration methods are used by the International GNSS Service. For the transmitters, only Galileo provides calibrations performed on ground before launch. It has been shown that this has an important impact on the observed scale factor. The GENESIS satellite, carrying a \ac{GNSS} receiver, can provide a lot of information about the scale and the motion of the geocenter (see Sect.~\ref{sec:GeocScale} for more details). 
Therefore, careful design of the GENESIS platform is required to minimize the sensitivity of the dynamic modeling to external effects such as direct and reflected solar radiation pressure. For the measurement of long-term variations of these accelerations, the currently used accelerometer technology, which requires a posteriori calibration, is not suitable.

Contributions to the performance of \ac{GNSS} by GENESIS will include the following: 

\begin{itemize}
    \item[1] The well-calibrated satellite platform will provide local tie vectors in space between physical antenna phase centres allowing for a high-precision linking of \ac{GNSS} observations with the other space geodetic techniques, referencing \ac{GNSS} positioning results to a well-defined reference frame (see Sect.~\ref{sec:RefFrames} and Sect.~\ref{sec:ITRSrealization})
    \item[2] The satellite antenna provides a clean absolute reference for accurate and consistent calibration of the transmitting antennas of all \ac{GNSS} satellites, without atmospheric propagation errors. In the case of Galileo, the antenna phase maps measured on the ground will allow a better validation of these calibrations  (see Sect.~\ref{sec:PCOPCM}). Pseudo-range biases will also be observed with much better accuracy due to the reduction of ionospheric pseudo-range errors (see Sect.~\ref{sec:IonoPlasma})
    \item[3] The specific observability in the radial direction of the satellite orbit allows \ac{GNSS} to independently contribute to the realization of the scale and origin of the Earth reference frame (see Sect.~\ref{sec:GeocScale}) 
\end{itemize}

As a proof of concept the capability of calibrating \ac{GPS} transmit antennas using receivers on board \ac{LEO} satellites without relying on an external scale was demonstrated by \citet{haines_realizing_2015} using \ac{GPS} tracking data from GRACE-B and TOPEX/Poseidon. 

\subsection{GNSS antenna phase centre calibration} \label{sec:PCOPCM}%
\label{sec:GNSS}

Conventionally a mechanical reference point is defined for each \ac{GNSS} antenna while the actual transmitting or receiving point might differ by up to few centimeters. For the receiving antenna these \ac{PCO}s and direction-dependent phase variations have been discussed since the early 1990's leading to sophisticated calibration methods \citep{rothacher1995, elosegui1995, mader1999, wubbena2000}. Compared deviations for the \ac{GNSS} satellites became apparent with the evolution of the \ac{GPS} constellation in the early 2000's \citep{zhu2003, ge2005, cardellach2007}. The problem was partially solved by estimating the \ac{GNSS} transmitter antenna patterns in the adjustment process \citep{schmid2003, barsever1998, schmid2005, dilssner2010, steigenberger2016}. This approach suffers from considerable limitations. The major restriction is the correlation between the terrestrial scale, the satellite clock, and the satellite antenna offsets resulting from the observation geometry. As one consequence the scale information is transferred from \ac{VLBI} and \ac{SLR} networks to the \ac{GNSS} network \citep{schmid2007} which prevents \ac{GNSS} from providing an independent scale. The second limit is given by the fact that the absolute antenna phase patterns of ground tracking sites are contaminated by local environmental effects such as time-variable multipath. A third limitation is given by the required estimation of tropospheric delays for each ground station. In 2016, Galileo released as first \ac{GNSS} ever precise calibrations for antenna phase center and phase variations. Applying them in the \ac{GNSS} estimation, differences between GPS- and Galileo-based coordinates become visible ~\citep{villigerAntennaCalibrationsTRF2018}. By fixing the Galileo \ac{PCO}s to the calibrated values, a Galileo-based scale is realized and the \ac{GPS} \ac{PCO}s were estimated simultaneously in an integrated processing ~\citep{villigerGNSSScaleDetermination2020}. 

A totally independent method to estimate scale-free \ac{GNSS} \ac{PCO}s is given via the usage of space-based \ac{GNSS} observations and the gravitational constraints from the orbital dynamics of the corresponding low Earth orbiter~\citep{huangEstimationGPSTransmitter2022}. The high consistency between the \ac{LEO}-based and the Galileo-based approaches has been shown by~\cite{huangTwoMethodsDetermine2021}. Despite the larger constellation of Galileo than that of \ac{LEO}s (24 versus 10+ in 2022), the \ac{LEO}-based approach has advantages in several important aspects. The additional geometry due to the fast movement and the altitudes of the \ac{LEO}s both benefit the de-correlation of the \ac{GPS} \ac{PCO}s and the scale. The altitudes of the \ac{LEO}s also lead to negligible impact of troposphere delay on the space-based observations. Moreover, the long-term available data of historical and operating \ac{LEO}s can be used for the estimation of the \ac{GPS} \ac{PCO}s backwards in time. Recent studies by~\cite{Glaser2020} and~\cite{huangEstimationGPSTransmitter2022} confirmed a one-millimeter accuracy requirement for the receiver \ac{PCO} position on board the LEOs. 

As the antennas of GENESIS will be fully calibrated together with the entire satellite structure, this mission offers a pure absolute reference for a precise and consistent calibration of the transmit antennas of all \ac{GNSS} satellites, including Galileo and BeiDou. The orbiting geodetic observatory thus offers the possibility to determine consistent \ac{GNSS} transmit phase patterns without relying on a scale from external sources, thus providing \ac{GNSS} with the capability to contribute independently to the realization of the scale of the terrestrial reference frame \citep{barsever2009, haines_realizing_2015, huangEstimationGPSTransmitter2022}. Additionally, the geometry between GENESIS and a \ac{GNSS} satellite allows scanning the transmitting antenna pattern within a short time and up to nadir angles of \ang{30}. Thus, it is possible to reduce additional correlations between the phase centre offsets and the spacecraft orientation. The extension of nadir angles to \ang{30} for transmitting antenna calibrations leads to a significant improvement in the \ac{GNSS}-based orbit determination for other scientific Earth observation missions. Therefore, GENESIS will allow improving products of such missions, especially, orbit dependent altimeter data.


\subsection{Positioning of satellites and space probes} %
\label{sec:PosSat}

GENESIS will provide a breakthrough in space geodesy and by this will contribute to improving the accuracy of many satellite orbits, and consequently, the accuracy of the parameters that these satellites observe. \ac{POD} is an integral part of analyzing the data of numerous Earth science missions and of inter-planetary space probes. Prominent examples of past, present and future missions devoted to Earth observation are radar altimetry missions such as TOPEX/Poseidon, Jason-1, -2, -3, Sentinel-3, and Sentinel-6A, laser altimetry missions such as ICESat-1 and -2, gravity missions such as CHAMP, GRACE, GOCE, and GRACE-FO, and many other missions such as the SAR/InSAR missions TerraSAR-X, TanDEM-X, and Sentinel-1, the magnetic field mission Swarm or further satellites of the Copernicus Earth observation program of the European Union. 

All above mentioned missions significantly rely on \ac{POD}, some of them even critically, as the quality of the science products may directly depend on the accuracy of the orbit determination, e.g., requiring a radial orbit accuracy of \SI{15}{mm} with a goal of \SI{10}{mm} for the science products of Sentinel-6A \citep{donlon_copernicus_2021} or a 3-D orbit accuracy of \SI{5}{cm} for the Sentinel-1 mission \citep{gmes_sentinel-1_team_gmes_2004}. This holds for all gravity missions and in particular for altimetry missions, where the radial component is of primary interest \citep{cerri_precision_2010}. Sea level measurements from radar altimetry, e.g., are directly related to this component \citep{abdalla_altimetry_2021}. For missions providing long-term climatological data records, it is therefore essential to perform the most accurate \ac{POD} in a reference frame which is consistent across many years for the data analysis of many different spacecraft. In addition to a stable reference frame, which is crucial to not contaminate sea level rise measurements with reference frame drifts \citep{altamimi_reference_2013}, the highly accurate modeling of non-gravitational \citep[e.g.,][]{flohrer_generating_2011, mao_dynamic_2021} and gravitational forces, e.g., the proper modeling of temporal gravity variations across many years \citep{couhert_towards_2015, peter_cost-g_2022}, as well as a full exploitation of multiple tracking techniques \citep{luthcke_1-centimeter_2003, choi_jason-1_2004} is mandatory for the most accurate \ac{POD}.

Similar to the orbit determination of the GENESIS spacecraft, \ac{POD} of any other Earth science spacecraft refers to the positioning of the satellite \ac{CoM} in a \ac{TRF}. \ac{POD} thereby occurs across a range of time-scales: near-real-time, intermediate latency, and longer latency for mission science products and climate data records. Missions with stringent accuracy requirements on \ac{POD} usually employ multiple and independent \ac{POD} payloads for this purpose. Onboard \ac{GNSS} receivers, \ac{DORIS} receivers, and \ac{SLR} reflectors are used to improve the quality and robustness of the orbit determination and to enable cross calibrations \citep[e.g.,][]{montenbruck_sentinel-6a_2021}. 

Usually \ac{GNSS} data provide one of the strongest \ac{POD} contributions due to the almost continuous tracking of all-in-view \ac{GNSS} satellites. As the quality of GNSS-based \ac{POD} critically depends on a proper modeling of systematic errors, e.g., phase centre variations of the \ac{GNSS} receiver and transmitter antennas \citep[see][Sect.~\ref{sec:PCOPCM}]{jaggi_phase_2009, schmid_absolute_2016}, the improved calibrations of the \ac{GNSS} transmitter antennas, that will be provided by GENESIS, will further improve the performance of \ac{GNSS}-based \ac{POD}. This is of particular relevance for \ac{GNSS} measurements collected at low elevation angles (or large nadir angles as seen from the \ac{GNSS}), where altimetry missions collect a large amount of data and where \ac{GNSS} transmitter calibrations are still poorly determined today \citep{schmid_absolute_2016}. The further reduction of systematic errors will pave the way towards mm-accurate \ac{GNSS} orbit determination of Earth science spacecraft, when exploiting the integer nature of \ac{GNSS} carrier phase ambiguities \citep{jaggi_precise_2007, bertiger_sub-centimeter_2010, montenbruck_precise_2018}. 

GENESIS will also benefit the \ac{DORIS}, \ac{SLR} and \ac{VLBI} techniques. For \ac{DORIS}, the limiting error source at present is the \ac{USO} sensitivity to radiation-induced perturbations, particularly while traversing the South Atlantic Anomaly \citep[][]{stepanek_inclusion_2020}. The possibility to synchronize the USO to atomic clocks on the ground thanks to an \ac{A-LRR} on board the GENESIS satellite would remove this source of error from the satellite segment. For \ac{SLR}, the data from the Jason-2/\ac{T2L2} experiment made it possible for the first time to globally calibrate the time biases in the stations of the \ac{SLR} network \citep{EXERTIER2017948}. The time-transfer experiment incorporated onto the GENESIS mission would be crucial to continue such a global calibration and will thus benefit global geodesy, and indirectly \ac{POD} by helping to improve the \ac{SLR} technique. Finally, the VLBI transmitter (VT, see Sect.~\ref{sec:VLBI}) on board GENESIS will provide the spacecraft position in the \ac{CRF}.

The GENESIS spacecraft will provide an additional platform to isolate residual biases in the ranging systems of \ac{SLR} stations \citep[e.g.,][]{luceri_systematic_2019}. These residual range errors are extremely challenging to isolate, because they are dependent on ranging  equipment and technique, the retro-reflector target response, the data sampling and editing when creating the \ac{SLR} normal points. GENESIS will provide a platform in a \ac{MEO} where the other tracking data (\ac{GNSS}, \ac{DORIS}, \ac{VLBI}) will help to isolate residual bias effects \citep[e.g.,][]{arnold_satellite_2019}.

Provided that the locations of the spacecraft \ac{CoM} and of the individual \ac{POD} sensors are all known in the respective satellite body-fixed frame with sufficient accuracy from pre-launch assessments, it will, therefore, be possible to translate the high precision of the \ac{GNSS}, \ac{DORIS}, and \ac{SLR} data collected by Earth science spacecraft into mm-accurate satellite positions expressed in the highly accurate and long-term stable reference frame provided by GENESIS. 

\subsection{Relativistic geodesy and time and frequency transfer} %
\label{sec:RelGeo}

The International Astronomical Union recommended to introduce the General Theory of Relativity as the theoretical background for the definition of space-time reference systems \citep{Soffel_2003}. Applying the General Theory of Relativity is indispensable to meet the required geodetic accuracy and stability of a \ac{TRF} -- i.e. \SI{1}{mm} for positions and \SI{0.1}{mm/year} for velocities -- for detecting smallest variations in the Earth system components. GENESIS will prove the consistent use of the General theory of Relativity in space geodesy and the involved reference systems at an unprecedented accuracy level.

The direct integration of gravimetric and geometric reference frames is an open issue, where time and frequency measurements will play a central role. For the realization of a dynamical reference such as a physical height system and the related equipotential surface, called the geoid, gravity field measurements are required. Today, such an equipotential surface is only known at the centimeter to decimeter level if comparing point values on larger scales. To overcome this shortcoming, highly precise optical clocks connected by dedicated ground or space links can be used. This novel technique has reached an accuracy that allows the precise measurement of differences of the gravity potential exploiting the gravitational redshift \citep{muller2018high, Delva2019}. Optical frequency standards at the leading national metrology institutes today show relative frequency inaccuracies in the $10^{-18}$ range (corresponding to \SI{1}{cm} in height) and beyond \citep{Brewer2019, bothwell_jila_2019, beloy_frequency_2021}. Long-distance optical frequency transfer using phase-stabilized optical fibers has been demonstrated with a relative frequency inaccuracy at the $10^{-19}$ level \citep{Lisdat2016}, and even $10^{-21}$ on shorter distances \citep{xu_reciprocity_2019}. Free-space laser links already realized Common View time transfer over thousands of kilometers at the picosecond level \citep[e.g., with \ac{T2L2}, see][]{EXERTIER2017948}, and reach on shorter distances the femtosecond level for time transfer \citep{sinclair_femtosecond_2019} and $10^{-21}$ level for frequency transfer \citep{gozzard_ultrastable_2022}. It is thus possible, with optical standards and frequency transfer techniques, to consistently unify height systems and to improve the accuracy of physical height reference frames and the geoid, to the cm level or better that is of prime interest in the context of \ac{GGOS} and, especially, for monitoring sea level change. GENESIS could enable a proof of principle for comparing physical heights (derived from time and frequency differences) over large distances if an \ac{A-LRR} will be present on board the satellite.

\section{Mission design and instruments}
\subsection{CDF study output}
\label{sec:CDF}
Using well-established \ac{ESA} \ac{CDF} assessment, the GENESIS Mission feasibility has been assessed during March/April 2022, with the contribution of over 50 \ac{ESA} experts covering all necessary mission feasibility expertise profiles.

The GENESIS high level Mission Objectives have been defined as follows:
\begin{itemize}
    \item \textbf{Obj-1:} Improve \ac{ITRF} accuracy and stability by providing in-orbit co-location and necessary combined processing of the four space geodetic techniques that contribute to its realization, namely \ac{GNSS}, \ac{SLR}, \ac{DORIS} and \ac{VLBI}, on a highly calibrated and stable platform.\\
    The goal is to contribute to the achievement of the \ac{GGOS} objectives for the \ac{ITRF} realisation, aiming for a parameter accuracy of \SI{1}{mm} and a stability of \SI{0.1}{mm/year} to the \ac{GGOS}, in order to provide significant scientific benefits in Earth modelling.
    
    \item \textbf{Obj-2:} To improve, compared to the current state of the art, the operational time and frequency transfer and synchronisation globally. Target performance \SI{10}{ps} in time transfer and $10^{-18}$ for relative frequency transfer. 
\end{itemize}
Thanks to the onboard \ac{VLBI}, which is the only geodetic technique allowing access to the \ac{ICRF}, GENESIS shall also allow obtaining a direct link between the \ac{ITRF} and the \ac{ICRF} (differencing the radio signal emitted by GENESIS with the signals from the fixed quasar radio sources).

The CDF study identified initial high-level Mission Requirements, given in Table~\ref{tab:requ}. From these, the following key technical drivers are identified:
\begin{enumerate}
    \item \textbf{The need of a Very Precise onboard Metrology (calibrated ties)}: the offset between each payload and the satellite \ac{CoM} shall be known with accuracies $< \SI{1}{mm}$. Offset stability shall remain within \SI{1}{mm}-level during the whole duration of the mission. (requiring adequate thermoelastic materials; extremely accurate on-ground calibration tests; etc).
    \item \textbf{A common time reference for all onboard instruments} shall be ensured (all geodetic instruments shall be referenced and synchronized to each other)
    \item \textbf{Highly accurate Precise Orbit Determination}: GENESIS shall be able to determine the orbit with accuracies at mm level ($< \SI{1}{cm}$) (excellent GNSS POD, requiring high success rate Integer ambiguity resolution and a very accurate radiation pressure model of the GENESIS satellite).
    \item \textbf{Simultaneous operation of Geodetic techniques}, guaranteeing the maximum contemporary use of the 4 onboard geodetic techniques (and at least 2 at all times) with best possible performances.   
\end{enumerate}

\begin{table*}[htb!]
    \caption{\label{tab:requ} First estimation of high-Level Mission Requirements for GENESIS.}
    \vspace{0.2 cm}
    \centering
    \begin{minipage}{1.\linewidth}
    \begin{tabularx}{1.0\textwidth}{p{2cm}p{16cm}}
\textbf{Req. ID} & \textbf{Statement} \\\hline\hline
001                          & The 				GENESIS mission shall be designed to achieve the main mission 				objective Obj-1, through the co-location in space of the 				following 4 geodetic techniques: GNSS, VLBI, SLR, DORIS.                                                                                                                                                                                                                                                        \\
\cellcolor[HTML]{DBE5F1}002  & \cellcolor[HTML]{DBE5F1}The 				GENESIS mission should be designed to achieve the main mission 				objective Obj-2, for operational time and frequency transfer and 				synchronisation                                                                                                                                                                                                                                                                        \\
003                          & The 				mission shall comply with the space debris mitigation regulations                                                                                                                                                                                                                                                                                                                                                                                         \\
\cellcolor[HTML]{DBE5F1}004  & \cellcolor[HTML]{DBE5F1}The 				casualty risk for the mission shall not exceed 1 in 10000 for 				any re-entry event (controlled or uncontrolled). If the predicted 				casualty risk for an uncontrolled re-entry exceeds this value, an 				uncontrolled re-entry is not allowed and a targeted controlled 				re-entry shall be performed in order not to exceed a risk level 				of 1 in 10000                                                              \\
005                          & The 				mission shall comply with the space debris mitigation 				requirements in the nominal and also in the failure case                                                                                                                                                                                                                                                                                                                                        \\
\cellcolor[HTML]{DBE5F1}006  & \cellcolor[HTML]{DBE5F1}The 				mission operational lifetime shall be at least 3 years, as 				a minimum, excluding LEOP, commissioning and disposal                                                                                                                                                                                                                                                                                                        \\
007                          & The 				GENESIS mission should be designed for a development time of 3 				to 4 years                                                                                                                                                                                                                                                                                                                                                                           \\
\cellcolor[HTML]{DBE5F1}008  & \cellcolor[HTML]{DBE5F1}The 				GENESIS mission should target a launch date in  2027                                                                                                                                                                                                                                                                                                                                                                      \\
009                          & Use 				of high-Technical Readiness Level, demonstrated instruments and payloads shall be 				preferred, whenever possible.                                                                                                                                                                                                                                                                                                                                                             \\
\cellcolor[HTML]{DBE5F1}010  & \cellcolor[HTML]{DBE5F1}Maximum 				re-use of existing facilities at \ac{ESA} and \ac{ESA} Member States                                                                                                                                                                                                                                                                                                                                                                    \\
011                          & The 				satellite should be launched into an orbit capable to fulfil the 				mission and payload requirements.                                                                                                                                                                                                                                                                                                                                                    \\
\cellcolor[HTML]{DBE5F1}012  & \cellcolor[HTML]{DBE5F1}A 				small satellite platform should be targeted                                                                                                                                                                                                                                                                                                                                                                                         \\
013                          & The 				satellite platform shall be able to accommodate all the GENESIS 				payloads associated to the geodetic techniques. Additionally, the 				platform should host the other enabling subsystems, as needed 				(and optional payloads as appropriate).                                                                                                                                                                                                     \\
\cellcolor[HTML]{DBE5F1}014  & \cellcolor[HTML]{DBE5F1}The 				platform nominal lifetime shall be at least 4 years.                                                                                                                                                                                                                                                                                                                                                                        \\
015                          & The 				offset between each payload and the satellite CoM shall be known 				with accuracy of 1 mm. Offset stability shall remain within \SI{1}{mm} level during the whole duration of the mission                                                                                                                                                                                                                                                                  \\
\cellcolor[HTML]{DBE5F1}016  & \cellcolor[HTML]{DBE5F1}The 				CoM position should be known with \SI{1}{mm} accuracy in the 				satellite reference frame.                                                                                                                                                                                                                                                                                                                                        \\
017                          & The 				satellite shall have a Nadir-pointing face for the whole mission 				duration, with a pointing accuracy less than 1 degree and a 				pointing stability of \SI{0.1}{degree} along the whole orbit.                                                                                                                                                                                                                                                                  \\
\cellcolor[HTML]{DBE5F1}018  & \cellcolor[HTML]{DBE5F1}The 				satellite platform shall be able to operate at the least 2 				geodetic techniques in parallel at all times.                                                                                                                                                                                                                                                                                                                      \\
019                          & Attitude 				determination shall be maintained at all times with accuracy 				below \SI{0.1}{degree}.                                                                                                                                                                                                                                                                                                                                                               \\
\cellcolor[HTML]{DBE5F1}020  & \cellcolor[HTML]{DBE5F1}The 				\ac{POD} will have to be able to determine the orbit with an accuracy 				better than \SI{1}{cm}. \ac{POD} is also affected by optical and thermal material properties (absorption, reflection and such) of the satellite outer surfaces to make an accurate radiation pressure model of the satellite. This has to be taken into account in \ac{CDF} in particular with respect to impact on costs.                                                                                                                                                                                                                                                                                                                                        \\
021                          & The S/C shall be able to download a volume of science data as follows:GNSS tracking data (1Hz) 0.2 GB/day; DORIS tracking data (0.1 Hz) 0.04 GB/day; Active SLR (SLR related, for synchronization between a ground clock linked to a laser station and a clock on board the 	satellite) \SI{0.05}{GB/day}; accelerometer \SI{0.002}{GB/day} \\
\cellcolor[HTML]{DBE5F1}022  & \cellcolor[HTML]{DBE5F1}To 				provide the link with current ITRF realizations, the selected 				orbit shall be accessible by the established global tracking 				networks of the different techniques                                                                                                                                                                                                                                                            \\
023                          & VLBI 				shall be visible for \SI{20}{\%} time from at the least 2 VLBI 				stations separated by \SI{10000}{km}                                                                                                                                                                                                                                                                                                                                                          \\
\cellcolor[HTML]{DBE5F1}024 & \cellcolor[HTML]{DBE5F1}The 				GENESIS Payloads are: GNSS receiver, VLBI transmitter, a DORIS 				Receiver and a SLR retroreflector.                                                                                                                                                                                                                                                                                                                           \\
025                         & A 				common time reference for all onboard instruments                                                                                                                                                                                                                                                                                                                                                                                                          \\
\cellcolor[HTML]{DBE5F1}026 & \cellcolor[HTML]{DBE5F1} An 				accelerometer for measuring the non-gravitational accelerations, 				contributing to POD should be included \\                                                                                                                                                                                                                                                               
\end{tabularx}
\end{minipage}

\end{table*}

As a result of the different optimisations performed at the \ac{CDF}, it has been concluded that a suitable compromise in terms of \ac{POD} and contribution of each technique results to be a \SI{6000}{km} circular orbit with quasi-polar inclination. This is selected as baseline orbit for GENESIS (ref. SS-01 from Table~\ref{tab:requ}).

The adoption of the circular orbit is compatible with requirement SS-11 from Table~\ref{tab:requ} and enables also the contemporary use of the 4 geodetic techniques at the same time and long baseline \ac{VLBI} observations (with more than \SI{6500}{km}) over \SI{75}{\%} of the time.

To reach this orbit two concepts have been conceived, within the program boundaries:
\begin{itemize}
    \item Satellite launched with VEGA-C in a sun-synchronous orbit in a piggyback configuration and the satellite being raised to the \SI{6000}{km} using electrical propulsion;
    \item Direct injection into \SI{6000}{km} orbit using a direct dedicated launch with several potential options being considered (e.g., Rocket Factory launcher, ISAR Aerospace launcher, a combined launch with Ariane 62/Galileo).
\end{itemize}
Depending on the launch option two different solutions have been defined with resulting satellite masses (wet) around \SI{375}{kg} (electrical propulsion option) and \SI{218}{kg} (Direct injection option).

\begin{figure}[t]
    \centering
    \includegraphics[width=\linewidth]{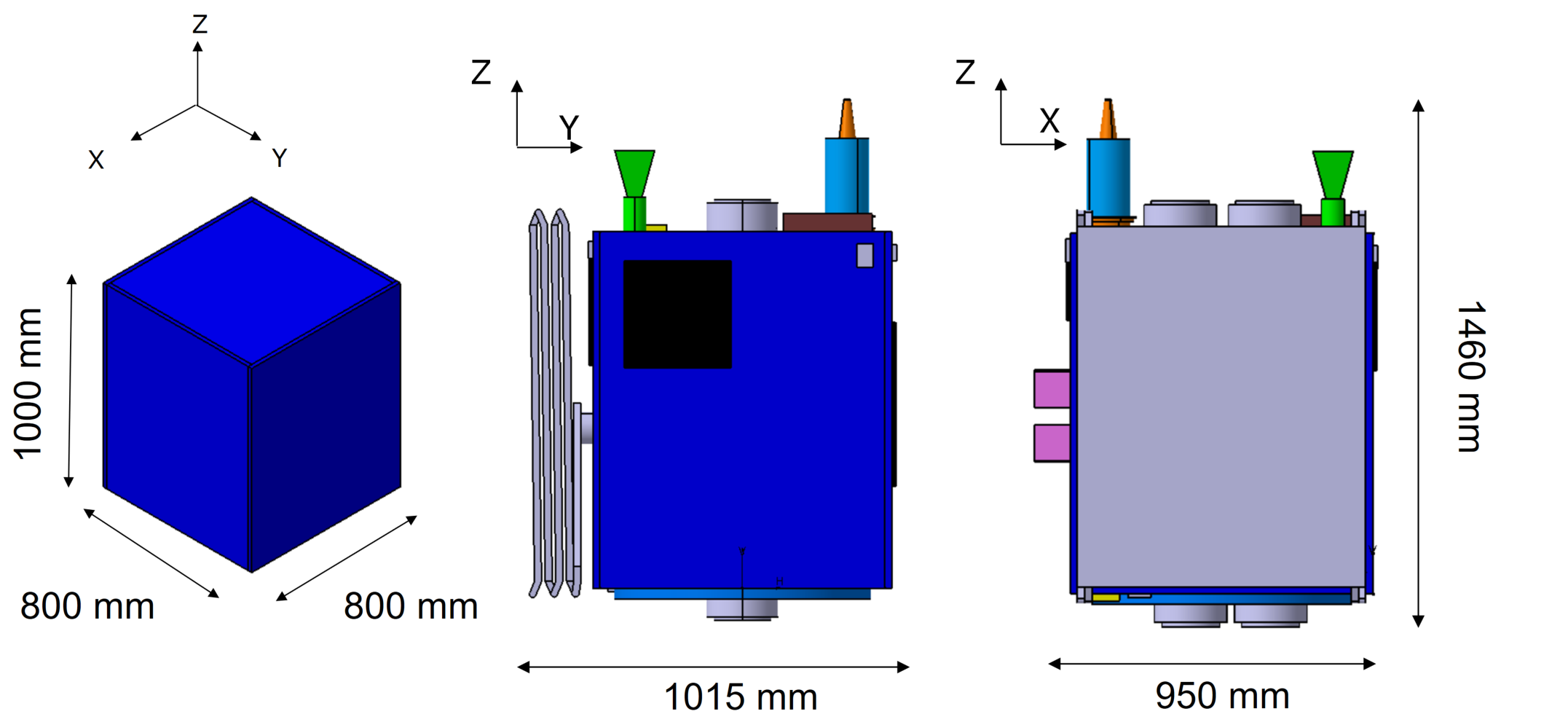}
    \caption{Satellite dimensions.}
    \label{fig:sat01}
\end{figure}

\begin{figure}[t]
    \centering
    \includegraphics[width=\linewidth]{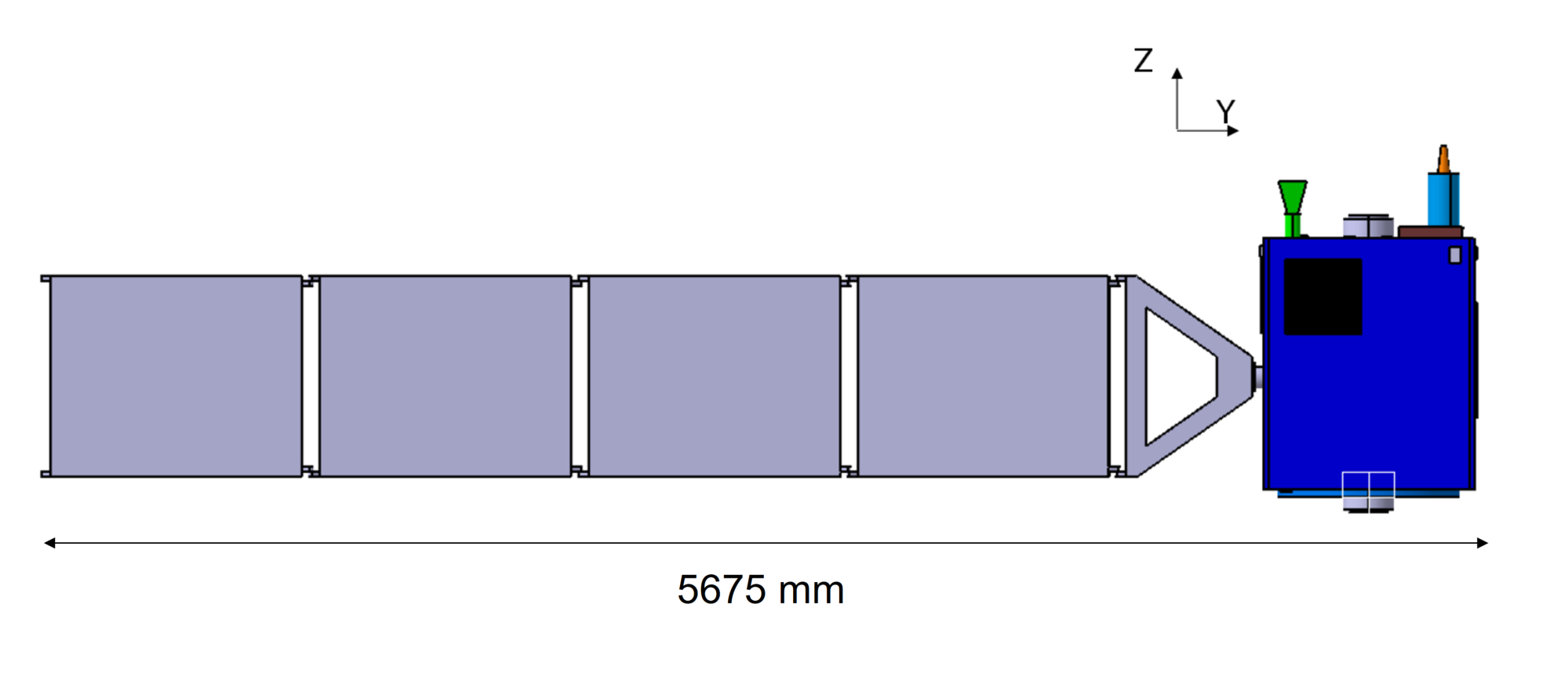}
    \caption{Solar panel dimensions.}
    \label{fig:sat02}
\end{figure}

The study has been instantiated for a generic platform, based on the PROBA-V concept, although it is concluded that several other platform options may be considered, resulting in the satellite dimensions shown in Figure~\ref{fig:sat01} and ~\ref{fig:sat02}.

The current realisation of the \ac{ITRF} is based on a multi-technique approach that suitably combines different observing methods taking advantage of their peculiar strengths. In particular, the frame origin is materialised by means of SLR observations; the frame scale is based on the average contribution of SLR and VLBI; DORIS disseminates the frame information to radar altimetry missions; and finally GNSS participates in a large number of ground ties, contributes to establishing and maintaining the conventional frame orientation and is essential to distribute the frame to a large community of users (see Sect.~\ref{sec:RefFrames0} for more details). Therefore, GENESIS will carry on board the payloads of all these space geodetic techniques, exploiting their co-location on a highly calibrated platform to further improve the ITRF accuracy and stability. The first ever co-location of VLBI with all satellite geodetic techniques will also strengthen the integration of Earth geometry and rotation. 

GENESIS will be equipped with an array of passive SLR retroreflectors (\ac{P-LRR}), a VLBI transmitter, one DORIS and a GNSS receiver (baseline configuration accounts for 2 GNSS Rx in cold redundancy connected to 2 sets of antennas on the nadir and zenith faces). The DORIS instrument and GNSS Rx will require specific adaptations to fly on a \SI{6000}{km} orbit. The additional installation of an \acf{A-LRR} device is also examined to allow accurate time transfer between SLR ground stations. At this stage, the \ac{A-LRR} is considered as an optional payload, but it is considered highly desirable and easy to be integrated as part of the baseline given its reduce mass/power requirements ($<\SI{1}{kg}$; $< \SI{1}{W}$). All active payloads will rely on a single time standard, realized by a \ac{USO} connected to a time distribution unit. An additional optional payload considered for GENESIS mission is the integration of an onboard accelerometer, to further support high-precision orbit determination, the modelling of non-conservative forces and allowing also in-orbit determination of the satellite \ac{CoM} (with several existing and under evolution instruments being identified in Europe).

\begin{figure}[b]
    \centering
    \includegraphics[width=\linewidth]{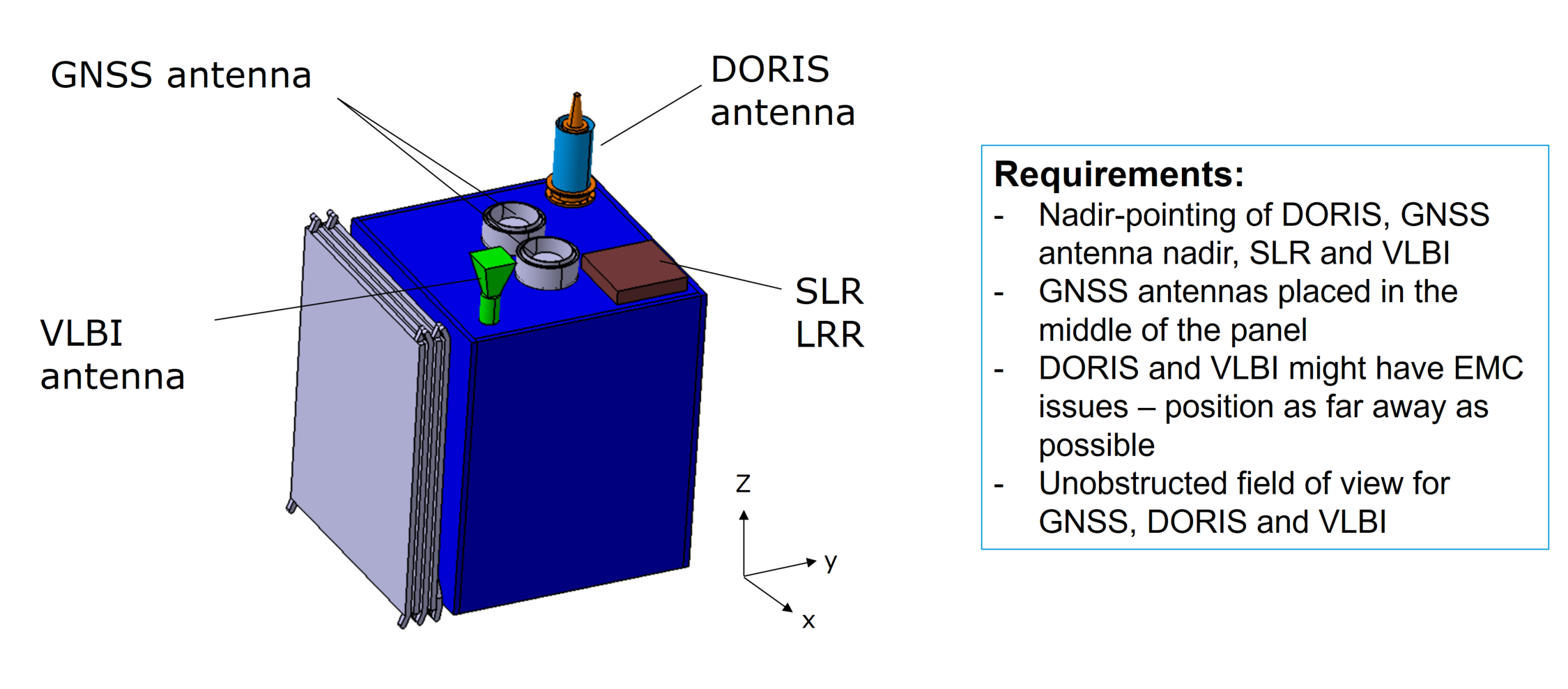}
    \caption{GENESIS instruments and requirements.}
    \label{fig:sat03}
\end{figure}

\begin{figure}[t]
    \centering
    \includegraphics[width=\linewidth]{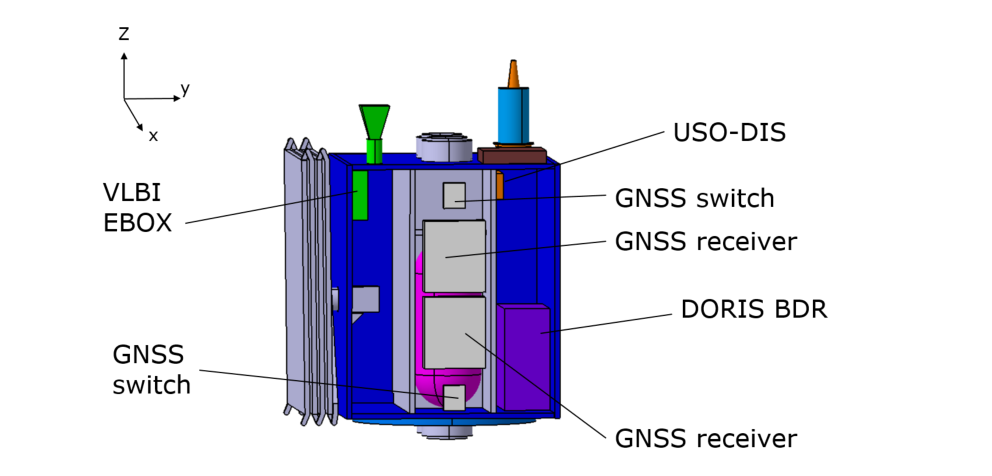}
    \caption{GENESIS instruments.}
    \label{fig:sat04}
\end{figure}

The design of the satellite and instruments integration is shown in Figure~\ref{fig:sat03} and~\ref{fig:sat04}.

In the framework of a potential international collaboration with NASA, an assessment has also been made on the possible integration of the NASA Geodetic Reference Instrument Transponder for Small Satellites (GRITTS) instrument. This instrument combines a GPS receiver and VLBI transmitter. The GRITSS concept is to upconvert the received GNSS signal and transponding it to VLBI stations (1-way biased range). This approach does not require the satellite to have a view of more than one VLBI station at a time, allowing it to be in \ac{LEO} but could be adapted to MEO orbit as well. If agreed to be included, it would increase the redundancy of geodetic payloads, and would support the simultaneous testing of different approaches for satellite VLBI.

The CDF GENESIS feasibility assessment has covered all GENESIS mission technical and programmatic aspects, including: system analysis; orbit options; mission analysis; chemical propulsion; electrical propulsion; AOCS; communications; data handling; power; thermal; structures; radiation; risks and programmatics; costs.

A detailed CDF GENESIS Report is currently under conclusion. A public version of this CDF Report will also be available. 

As a  main result of the \ac{ESA} \ac{CDF} assessment, the GENESIS Mission has been confirmed to be feasible within the \ac{ESA} FutureNAV defined program Boundaries, with a target launch date in 2027 (assuming the program is started in Q1 2023).

\subsection{Passive and active Laser Retro-Reflector} %
\label{sec:laser}

The \ac{SLR} observable is the round-trip time of flight of a laser pulse between a ground station and a target equipped with a Laser Retro-Reflector. Timing and time transfer techniques are at the heart of this activity. This technique makes a fundamental contribution to the establishment of the \ac{ITRF}, thanks to a network of observatories spread over a large part of the globe, the International Laser Ranging Satellite Network (see Sect.~\ref{sec:RefFrames0} for more details). The majority of satellites tracked by the \ac{ILRS} carry \ac{P-LRR}. The size, the number and the arrangement of Corner Cube laser Retro-reflector are a compromise between the link budget at the given satellite altitude, the orbit eccentricity and the metrological performances. 

Corner Cube laser Retro-reflector can be arranged spherically, e.g., in LAGEOS, LAGEOS-2 and LARES-2 (which are premiere geodetic satellites of the \ac{ILRS}) or as a flat panel as for the Galileo satellites \citep{url-ILRS}. Spherical arrays give the best accuracies. For example, LARES-2 (\SI{40}{\cm} diameter, $\sim \SI{300}{kg}$ mass) is a \ac{P-LRR} delivered by the Italian Space Agency designed to achieve a \ac{SLR} accuracy of \SI{1}{mm} (compared to the \SI{5}{mm} of LAGEOS). It was launched successfully by ESA on July 13, 2022, with the qualification flight of the Vega C and it was successfully tracked by the \ac{ILRS} \ac{SLR} network from July 14. Another spherical \ac{P-LRR} of significantly reduced size ($\sim \SI{10}{cm}$) and mass (kg level) compared to LARES-2 has been designed to achieve a SLR accuracy of \SI{2}{mm} for LEO satellites to be launched from 2024.

Alternatively, an \ac{A-LRR} can synchronise ground-based atomic clocks at intercontinental distances using standard satellite laser ranging techniques. An \ac{A-LRR} allows the precise determination of the onboard clock and the monitoring of its behaviour in the space environment (gravity field, radiation, temperature) as a supplement to the \ac{POD} provided by the conventional \ac{SLR}. Such an idea has been demonstrated by \ac{SLR} ground stations with the \ac{T2L2} instrument on board the Jason-2 satellite \citep{belli_temperature_2016}. This project highlighted the wide disparity between laser stations in local time management. This demonstration of time transfer by laser link is also expected for the European Laser Timing instrument of the \ac{ACES} experiment on board the International Space Station, which should be launched in 2024 \citep{cacciapuoti_testing_2020}.

An \ac{A-LRR} on board GENESIS would allow:
\begin{itemize}
\item to benefit from a modern retro-reflector designed to achieve millimeter accuracy and metrologically attached to the other space geodetic instruments;
\item be able to perform ground-to-space and ground-to-ground frequency and time transfers with an extended common view compared to \ac{T2L2} and \ac{ACES} missions by taking advantage of the higher altitude of the satellite;
\item to compare \ac{GNSS} and laser time transfer techniques with an uncertainty below \SI{100}{\ps};
\item to be able to accurately monitor the behaviour of the onboard clock for precise orbitography. 
\end{itemize}

The \ac{P-LRR} is typically less expensive than an \ac{A-LRR} counterpart. Being completely passive, is does not use any resources on board GENESIS (except its mass and volume envelope). A combination of both a \ac{P-LRR} and \ac{A-LRR} would provide all combined benefits, as well as the sum of the resources and mass/volume envelopes needed on board GENESIS.

\subsection{VLBI Transmitter} %
\label{sec:VLBI}


Observations to distant radio sources, such as quasars, with the \ac{VLBI} technique enable the determination of a space-fixed celestial reference frame like the \ac{ICRF} with its current realization ICRF3 \citep{Charlot2020} and the \acf{EOP}. \ac{VLBI} is the only technique that provides the full set of \ac{EOP} (polar motion, UT1, celestial pole offsets), contributes to the \ac{ITRF} scale, and uniquely realizes \ac{ICRF}, as opposed to the satellite-based techniques (see Sect.~\ref{sec:GeocScale}). The connection of the \ac{ITRF} to the \ac{ICRF} through \ac{VLBI} enables the study of the dynamics of the interior of the Earth through the wandering of the motion of the poles with respect to the celestial frame, as well as studying tidal dissipation and seasonal or interannual effects in the geophysical fluids on the solid Earth through measurements of the rotation rate of the Earth (see Sect.~\ref{sec:UnifFrame}). 

The fundamental \ac{VLBI} measurements are signal delay observations of the incoming radio signals between pairs of tracking stations. Over time, and using multiple distant radio sources, these measurements enable the determination of the baseline vectors between pairs of stations. The orientation of these vectors in the \ac{CRF} defines the Earth orientation in that frame. The lengths of these vectors, known to millimeter accuracy, critically contribute to the determination of the scale of the \ac{ITRF} (\citealp{altamimi2016}). The \ac{VLBI} observation campaigns and data processing are coordinated by the International \ac{VLBI} Service for Geodesy and Astrometry \citep{url-IVS}. 


One of the unique features of GENESIS is the \ac{VT} that will accurately connect geodetic \ac{VLBI} stations through a space-tie to the other geodetic techniques. The \ac{VT} instrument will transmit signals in different frequency bands in order to eliminate the ionospheric dispersive delay along the paths to each observing \ac{VLBI} station, and comply with the evolving observations procedures at all \ac{VLBI} stations. The signals can be observed by all geodetic \ac{VLBI} stations, including the new \ac{VGOS} fast slewing stations that are coming online, in their standard geodetic receiver setups. The \ac{VT} will exploit the full extent of the frequency bands allocated on a worldwide co-primary basis to the Earth exploration satellite service through the International Telecommunications Union Radio-communication Sector. The ultra-low-power density signals of the \ac{VT} will be well below the applicable coordination thresholds, ensuring easy compatibility with ITU rules.  Using the conventional \ac{VLBI} technique of correlating received signals across baselines it will be possible not only to determine the baseline vectors, as in conventional \ac{VLBI}, but also the absolute geocentric position of the receiving sites. \ac{VLBI} observations of GENESIS will therefore enable these stations to be accurately located within the GENESIS \ac{TRF} consistently with the other geodetic techniques, enable a frame tie between the celestial frame and the dynamic reference frames of satellite orbits as well as a frame tie between the GENESIS \ac{TRF} and the extremely accurate and stable inertial celestial frame. 

A European \ac{VT}, compatible with the accommodation constraints on board the Galileo satellite, performance of the receiving stations as well as with the ITU regulations in all transmission frequency bands is currently under development for consideration of Galileo second generation satellites \citep{url-ESA}. The \ac{VT} is currently designed to transmit at different frequencies between \SI{2}{GHz} to \SI{14}{GHz} but also higher frequency bands can be considered. The present setup for regular \ac{VGOS} observations use four \SI{1}{GHz} wide bands within the S, C, and X frequency bands. Discussions of including higher frequencies in Ka band is ongoing in order to maximise the covered signal bandwidth since the \ac{VLBI} estimates of group delay is approximated by the reciprocal of the observed bandwidth. It may be possible to investigate linking the \ac{ICRF} across frequency bands using the GENESIS satellite as a well-calibrated multi-frequency target. The \ac{VT} is designed to transmit both pseudo-noise and random noise. 
The random noise signal mimics the broader-band noise emitted by quasar radio sources routinely observed by \ac{VLBI}, hence can be processed by essentially the usual station software. The required and projected precision of phase measurement supported by the \ac{VT} are \SI{0.1}{mm} and \SI{0.01}{mm}, respectively, for 1 second observable (see \citealp{biancale_e-grasp_2017}).

The feasibility of a dedicated \ac{VLBI} transmitter on board future Galileo satellites and the assessment of the impact of quality and quantity of satellite observations on the derived geodetic parameters were studied recently by various authors (see \citealp{jaradat2021}, \citealp{klopotek_geodetic_2020}, \citealp{sert2022} and references therein). The concept of measurement and equipment choices were discussed in \citet{jaradat2021} and the signals designed to mimic the quasars' radiation as observed and recorded by ground-based telescopes were simulated. The output signal of this chain using \SI{2}{GHz} to \SI{11}{GHz} for VGOS, was also tested using a \ac{VLBI} baseband data simulator, then correlated and fringe-fitted for validation. 
It was shown that the combination of quasar and satellite observations could allow theoretically for simultaneous estimation of Earth Rotation Parameters (polar motion and UT1-UTC) along with geocenter offsets, \ac{VLBI} station positions and satellite orbits \citep{klopotek_geodetic_2020}. In the case of carefully selected satellite observations and optimized scheduling with the VGOS-type network, detection of geocenter motion could be feasible. The use of \ac{VT} on \ac{GNSS} satellites and observation assessment including the UT1-UTC transfer quality for Galileo orbits was demonstrated in \citet{sert2022}. These recent studies confirm the significance of dedicated onboard transmitters and the potential of geodetic \ac{VLBI} as another space geodetic technique.

Observations of quasars and GENESIS satellite within the same sessions will provide the geodetic community with a great opportunity for directly linking the dynamical reference frame of satellite orbits to the quasi-inertial reference frame of extra-galactic radio sources and redefining the role of \ac{VLBI} in space geodesy.

\section{Conclusion}
The first objective of the GENESIS mission is to contribute to the achievement of the GGOS accuracy and stability goals concerning the \ac{ITRS} realization, aiming for \SI{1}{mm} and \SI{0.1}{mm/year}, respectively. To this aim, GENESIS will provide in-orbit co-location of the four space geodetic techniques: \ac{GNSS}, \ac{SLR}, \ac{DORIS}, and \ac{VLBI}, on a highly calibrated and stable platform. 

\begin{figure}
    \centering
    \includegraphics[width=\linewidth]{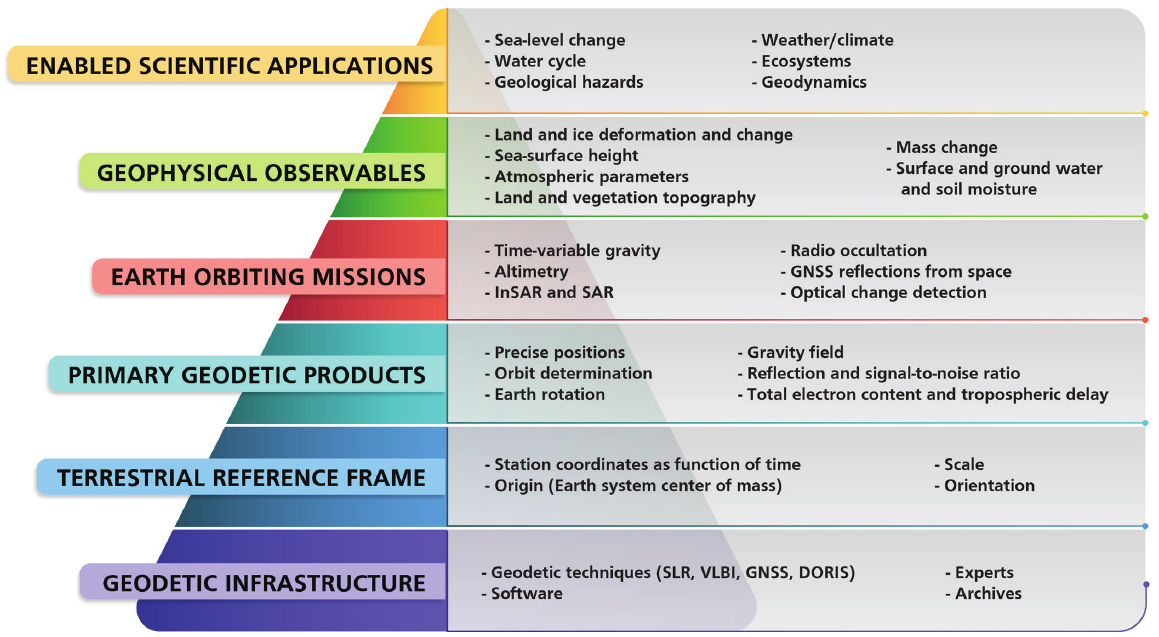}
    \caption{Illustration of how the geodetic infrastructure is linked to enabled scientific applications. \citep{national_academies_of_sciences_evolving_2020}}
    \label{fig:triangle}
\end{figure}

We have shown in this article the primary and critical importance of the \acf{ITRF} and associated geodetic infrastructure and products for many scientific applications in Earth and navigation sciences. This is illustrated in Figure~\ref{fig:triangle}. In particular, the accuracy and stability of the \ac{ITRF} is very important in the context of climate change to measure sea-level rise, improve estimates of ice mass balance, and determine Earth's energy imbalance -- which are all observables that are critical in climate change studies. Moreover, the \ac{ITRF} improvements can affect and improve many geodetic and geophysical observables as well as precise navigation and positioning. 

In addition, the \ac{ITRF} improvements strengthen the geodetic infrastructure, including the Galileo constellation, by reducing biases and errors between different techniques. GENESIS will be complementary and enhance the products of several other missions such as gravimetry and altimetry satellites. The addition of optional payloads such as an \acf{A-LRR} and an accelerometer can pave the way for very interesting and complementary objectives in navigation and time/frequency metrology.

The crucial necessity of an accurate and stable realization of \ac{ITRS} is endorsed by a large community of scientists and industries as well as various authorities, including the \ac{IAG} and the \ac{UN}. Finally, a study by the \ac{CDF} of \ac{ESA} has demonstrated the feasibility of the GENESIS mission within the \ac{ESA} FutureNAV defined program boundaries, with a targeted launch date in 2027.

\section*{Acronyms used}  
\begin{acronym}[GRACE-FO] 

\acro{ACES}{Atomic Clock Ensemble in Space}
\acro{A-LRR}{Active Laser Retro-Reflector}

\acro{CDF}{Concurrent Design Facility}
\acro{CERES}{Clouds and the Earth's Radiant Energy System}
\acro{CF}{Center of Figure\acroextra{ (of the Earth)}}
\acro{CM}{Center of Mass\acroextra{ (of the Earth)}}
\acro{CoM}{Center of Mass\acroextra{ (of the satellite)}}
\acro{CRF}{Celestial Reference Frame}

\acro{DGFI-TUM}{Deutsches Geod{\"a}tisches Forschungsinstitut at Technische Universit{\"a}t M{\"u}nchen}
\acro{DORIS}{Doppler Orbitography and Radiopositioning Integrated by Satellite}

\acro{EEI}{Earth Energy Imbalance}
\acro{EOP}{Earth Orientation Parameters}
\acro{ESA}{European Space Agency}

\acro{GGOS}{Global Geodetic Observing System}
\acro{GGRF}{Global Geodetic Reference Frame}
\acro{GIA}{Glacial Isostatic Adjustment}
\acro{GMSL}{Global Mean Sea Level}
\acro{GNSS}{Global Navigation Satellite System}
\acro{GPS}{Global Positioning System}
\acro{GRACE}{Gravity Recovery and Climate Experiment}
\acro{GRACE-FO}{Gravity Recovery and Climate Experiment Follow On}

\acro{IAG}{International Association of Geodesy}
\acro{ICESat}{Ice, Cloud, and land Elevation Satellite}
\acro{ICESat-2}{Ice, Cloud, and land Elevation Satellite 2}
\acro{ICRF}{International Celestial Reference Frame}
\acro{ICRS}{International Celestial Reference System}
\acro{IERS}{International Earth Rotation and Reference Systems Service}
\acro{IGN}{Institut national de l'information g{\'e}ographique et foresti{\`e}re}
\acro{ILRS}{International Laser Ranging Service}
\acro{IPCC}{Intergovernmental Panel on Climate Change}
\acro{ITRF}{International Terrestrial Reference Frame}
\acro{ITRS}{International Terrestrial Reference System}
\acro{IUGG}{International Union of Geodesy and Geophysics}

\acro{JPL}{Jet Propulsion Laboratory}

\acro{LAGEOS}{LAser GEOdynamics Satellites}
\acro{LARES}{LAser RElativity Satellite}
\acro{LEO}{Low-Earth Orbit}
\acro{LLR}{Lunar Laser Ranging}

\acro{MEO}{Medium-Earth Orbit}

\acro{OHU}{Ocean Heat Uptake}

\acro{PCO}{Phase Center Offsets}
\acro{PV}{Phase Variations}
\acro{P-LRR}{Passive Laser Retro-Reflector}
\acro{POD}{Precise Orbit Determination}

\acro{SLR}{Satellite Laser Ranging}
\acro{SRP}{Solar Radiation Pressure}
\acro{SDGs}{Sustainable Development Goals}

\acro{T2L2}{Time Transfer by Laser Link}
\acro{TEC}{Total Electron Content}
\acro{TOA}{Top Of the Atmosphere}
\acro{TRF}{Terrestrial Reference Frame}
\acro{TRS}{Terrestrial Reference System}

\acro{UN}{United Nations}
\acro{UN-GGIM}{United Nations Committee of Experts on Global Geospatial Information Management}

\acro{VGOS}{VLBI Global Observing System}
\acro{VLBI}{Very Long Baseline Interferometry}
\acro{VT}{VLBI Transmitter}

\acro{USO}{Ultra-Stable Oscillator}

\end{acronym}

\section*{acknowledgments} JB and AN are grateful to the Austrian Science Fund (FWF) for supporting this work with project P33925. Work by SD has been supported by INFN and by ASI (Italian Space Agency) under the ASI-INFN Joint Lab Agreement n. 2019-15-HH.0. The contribution of JM was supported by the Deutsche Forschungsgemeinschaft (DFG, German Research Foundation) via Collaborative Research Center CRC 1464 “TerraQ”, project-ID 434617780, and Germany’s Excellence Strategy EXC 2123 “QuantumFrontiers”, project-ID 390837967. Work by SG has been supported by the German Research Foundation (DFG) under Grant Number SCHU 1103/8-1 (GGOS-SIM, Simulation of the Global Geodetic Observing System) and SCHU 1103/8-2 (GGOS-SIM-2). 

The GENESIS mission is supported by many scientists, industrial partners, and space agencies, namely: Elisa Felicitas Arias (Paris Observatory-PSL, France), François Barlier (C{\^o}te d'Azur Observatory, France), Bruno Bertrand (Royal Observatory of Belgium, Belgium), Claude Boucher (Bureau des Longitudes, France), Sara Bruni (PosiTim UG at ESA/ESOC, Germany), Carine Bruyninx (Royal Observatory of Belgium, Belgium), Hugues Capdeville (CLS, France), Corentin Caudron (Université libre de Bruxelles, Belgium), Julien Chabé (C{\^o}te d'Azur Observatory, France), Sara Consorti (Thales Alenia Space, Italy), Christophe Craeye (Universit{\'e} catholique de Louvain, Belgium), Pascale Defraigne (Royal Observatory of Belgium, Belgium), Clovis De Matos (ESA/HQ, France), Jan Dou{\u s}a (Geodetic Observatory Pecny, Czech Republic), Fabio Dovis (Politecnico di Torino, Italy), Frank Flechtner (GFZ German Research Centre for Geosciences, Potsdam, Germany), Claudia Flohrer (BKG, Germany), Aurélien Hees (Paris Observatory-PSL/CNRS, France), René Jr. Landry (Qu{\'e}bec University, Canada), Juliette Legrand (Royal Observatory of Belgium, Belgium), Jean-Michel Lemoine (GET/CNES/CNRS, France), David Lucchesi (IAPS/INAF, Italy), Marco Lucente (IAPS/INAF, Italy), Nijat Mammadaliyev (Technische Universität Berlin, GFZ Potsdam, Germany), Gr{\'e}goire Martinot-Lagarde (C{\^o}te d'Azur Observatory, France), Stephen Merkowitz (NASA GFSC, United States), Gaetano Mileti (University of Neuch{\^a}tel, Switzerland)
, Terry Moore (University of Nottingham, United Kingdom), Juraj Papco (Slovak University of Technology, Slovakia)
, Roberto Peron (IAPS/INAF, Italy), Paul Rebischung (IGN/IPGP, France), Pascal Rosenblatt, LPG/CNRS (France), Séverine Rosat, ITES-EOST/CNRS (France), Matteo Luca Ruggiero, Università degli Studi di Torino (Italy), Alvaro Santamaria (Universit{\'e} Paul Sabatier, France), Francesco Santoli ( IAPS/INAF, Italy), Feliciana Sapio ( IAPS/INAF, Italy), Jaume Sanz (Universitat Polit{\`e}cnica de Catalunya, Spain), Patrick Schreiner (GFZ German Research Centre for Geosciences, Potsdam, Germany), Erik Schoenemann (ESA/ESOC, Germany), Harald Schuh (GFZ German Research Centre for Geosciences, Potsdam, Germany), Laurent Soudarin (CLS, France), Cosimo Stallo (Thales Alenia Space, Italy), Dariusz Strugarek (Wroclaw University of Environmental and Life Sciences, Poland), Angelo Tartaglia (INAF, Italy), Daniela Thaller (BKG, Germany), Maarten Vergauwen (KU Leuven, Belgium), Francesco Vespe (Agenzia Spaziale Italiana, Italy), Massimo Visco (IAPS/INAF, Italy), Jens Wickert (GFZ German Research Centre for Geosciences, Potsdam, Germany) and Pawe{\l} Wielgosz (University of Warmia and Mazury, Poland).

\bibliographystyle{apalike}
\bibliography{main}

\end{document}